\documentclass[12pt]{article}
 
\usepackage{hyperref}
\usepackage{rotfloat}
\usepackage{array} 
\usepackage{amssymb}
\usepackage{graphics,graphpap}
\usepackage{graphicx}
\usepackage{color}
\usepackage{graphicx}
\usepackage{dcolumn}
\usepackage{epsfig}
\usepackage{epstopdf}
\usepackage{bm}
\usepackage{amsmath}
\usepackage{amsfonts}
\usepackage{textcomp}
\usepackage{bbm}
\usepackage{setspace}
\RequirePackage[normalem]{ulem} 

\renewcommand{\thefootnote}{\fnsymbol{footnote}}

\renewcommand{\Im}{\mathrm{Im}}

\def\simge{\mathrel{%
   \rlap{\raise 0.511ex \hbox{$>$}}{\lower 0.511ex \hbox{$\sim$}}}}

\def\simle{\mathrel{
   \rlap{\raise 0.511ex \hbox{$<$}}{\lower 0.511ex \hbox{$\sim$}}}}

\def\s#1{\setbox0=\hbox{$#1$}%
\rlap{\ifdim\wd0>.7em\kern.22\wd0\else\kern.1\wd0\fi /}#1}

\newcommand{\aver}[1]{\langle #1\rangle}

\newcommand{\al}{\alpha}

\newcommand{\GeV}{\,{\rm GeV}}

\newcommand{\matel}[3]{\langle #1|#2|#3\rangle}

\newcommand{\mi}{\!-\!}
\newcommand{\C}{\mathbb{C}}
\newcommand{\ReL}{{\rm Re}}
\newcommand{\ImL}{{\rm Im}}

\begin{document}

\setcounter{footnote}{0}
\renewcommand{\thefootnote}{\arabic{footnote}}

\begin{titlepage}
\begin{flushright}\begin{tabular}{l}
Edinburgh/12/17 \\
CP$^3$-Origins-2012-29 \\
DIAS-2012-30
\end{tabular}
\end{flushright}
\vskip1.5cm
\begin{center}
   {\large \bf \boldmath Exclusive Chromomagnetism in  heavy-to-light  FCNCs}
    \vskip1.3cm {\sc
Maria Dimou$^{b,}$\footnote{md1e10@soton.ac.uk}, James Lyon$^{a,b,}\footnote{  J.D.Lyon@sms.ed.ac.uk} $ \& Roman Zwicky$^{a,b,}$\footnote{Roman.Zwicky@ed.ac.uk}
  \vskip0.5cm
        {\em  $^a$ {\sl School of Physics \& Astronomy, University of Edinburgh, 
    Edinburgh EH9 3JZ, Scotland} \\
  $^b$ {\sl School of Physics \& Astronomy, University of Southampton, 
    Highfield, Southampton SO17 1BJ, UK} \,  }} \\

\vskip2cm

{\large\bf Abstract:\\[8pt]} \parbox[t]{\textwidth}{
We compute matrix elements of the chromomagnetic operator, often denoted by  $ {\cal O}_8$, between 
$B/D$-states and light mesons plus an off-shell photon by employing the method of  light-cone sum rules (LCSR) at leading twist-2.
These matrix elements are relevant for processes such as $B \to K^* l^+l^-$ and they 
can be seen as the analogues of the well-known penguin form factors $T_{1,2,3}$ and $f_T$.  
We find a large CP-even phase for which we give a long-distance (LD) interpretation.
We  compare our results to QCD factorisation for which the spectator photon 
emission is end-point divergent.
The analytic structure of the correlation function used in our method admits a complex anomalous threshold 
 on  the physical sheet. The meaning and handling within the sum rule approach 
of the anomalous threshold  is discussed. 

}

\vfill

\end{center}
\end{titlepage}

\setcounter{footnote}{0}
\renewcommand{\thefootnote}{\arabic{footnote}}

\tableofcontents

\newpage

\section{Introduction}

Using  the method of LCSR, c.f. \cite{CK00} for a review,  we present a computation of transition matrix elements:
\begin{equation}
\label{eq:schematic}
\matel{ M(p) \gamma^*(q)}{{\cal O}_8}{H(p_H)} \;, \qquad p_H = p+q \;,
\end{equation}
of the chromomagnetic operator\footnote{
Our normalisation of $\mathcal O_8$ goes with the effective Hamiltonian normalisation convention: $\mathcal H_{\text{eff}}=-G_FV^*_{ts}V_{tb} C_8\mathcal O_8/\sqrt 2+\dots$}:
\begin{equation}
\label{eq:O8}
 {\cal O}_8  \equiv - \frac{g}{8 \pi^2} \, m_b    \bar s \sigma_{\mu\nu}  G_a^{\mu\nu}\frac{\lambda^a}{2}  (1+\gamma_5) b   \equiv  \left[ - \frac{g m_b}{8 \pi^2} \right] \, \tilde {\cal O}_8 \;,
 \end{equation}
from the lowest lying meson $J^P = 0^-$, denoted by $H$, with  with one heavy (beauty/charm) quark  to a light pseudoscalar(vector) meson $M$ and a photon. 
Allowing the latter to be off-shell, leads to photon momentum invariant $q^2$-dependence of the matrix element\footnote{We refrain from calling  these matrix elements  form factors since  they entail LD contributions  leading to a strong (CP-even) phase.}.
To  our knowledge this work represents the first computation of the matrix elements \eqref{eq:schematic}.
Factorisable parts have been computed in \cite{BB01,BFS01} to leading order in $1/m_b$ and next leading order, though with endpoint divergences \cite{KN01,FM02}, in QCD factorisation (QCDF)
as well as perturbative QCD (pQCD) \cite{pQCD}.
 For the $B \to K l^+l^-$ transition 
the 3-particle $B$-meson state has been computed in LCSR recently
\cite{Khodjamirian:2012rm}.

We find that the matrix elements are  suppressed by one(two) orders of magnitude 
for the $D(B)$-transitions w.r.t. to the penguin short-distance (SD) form factors. Their interest is thus
 for asymmetries rather than for branching ratios.
One example is  the isospin asymmetry since the emission of the photon from the spectator quark 
is dependent on the charge of the decaying hadron; another observable is CP-violation, 
in combination with new weak phases, where the strong phase leads to direct CP-violation \cite{IK12,LZ12a}. 
 We shall also dwell on the nature of the endpoint divergences found in QCDF
and how they relate to LCSR results in which they are absent.


The paper is organised as follows: In section \ref{sec:FF} we define the matrix elements and present the basic 
sum rule including a brief discussion  on anomalous thresholds and dispersion relations.
In section \ref{sec:computation} the computation is presented including the final sum rule expression. In section \ref{sec:numerics} the numerics for the matrix elements 
are detailed as well as qualitative discussions. 
In section \ref{sec:QCDF} we compare our results with the QCDF computation
in regard to endpoint divergences. In section \ref{sec:conclusions} we summarise the main points of the 
paper. Some explicit results and definitions can be found in appendices \ref{app:results} to \ref{app:DA}.
Ward-Takahashi identities, clarifying the r\^ole of contact terms and the analytic structure of the correlation 
functions in use, can be found in appendices \ref{app:WTI} and \ref{app:dispersion} respectively. 
A shorter write-up of some of the main points of this paper can be found in \cite{capri}.

\section{ Matrix element and sum rule}
\label{sec:FF}
\subsection{Lorentz-decomposition of $\tilde {\cal O}_8$ matrix elements}
\label{sec:defFF}

For definiteness, throughout this work, we shall choose the initial state meson to be of the  $\bar B$ type
and the final state meson to be a vector meson $V$. Replacements of $B$ by $D$-mesons and vector $V$ by pseudoscalar $P$  are self-understood. 
The amplitude of the  chromomagnetic operator, with uncontracted photon polarization vector $\epsilon(q)_\rho$  reads\footnote{Note the right hand side (RHS) of Eq.~\eqref{eq:matel} should be taken as a definition of the
matrix element  ${\cal A}^{* \rho}(V)$ in the case where the photon is off-shell.}:
\begin{eqnarray}
\label{eq:matel}
 {\cal A}^{* \rho}(V)   &\equiv& \matel{\gamma^*(q,\rho) V(p,\eta)}{\tilde O_8}{\bar B(p_B)} 
= i  \int_x \matel{V}{T j^\rho_{\rm em}(x)\tilde {\cal  O}_8(0)}{\bar B} e^{i q \cdot x}  + \dots \;.
\end{eqnarray}
The dots stand for higher-twist  photon distribution amplitude (DA) contributions. 
The former are neglected whereas the latter 
are briefly discuss in appendix \ref{app:photonDA}.
The polarisation vector of $V$ is denoted by $\eta$ and the momenta of $V$, $\gamma$ and $B$ 
are denoted   $p$, $q$ and $p_B \equiv p+q$  respectively. 
Here and thereafter we use: $\int_x = \int d^4 x$.
The star indicates that the photon is, generically,  off-shell.
The operator $\tilde {\cal O}_8 \equiv \bar s \sigma \cdot G (1+\gamma_5) b$ corresponds to 
${\cal O}_8$ \eqref{eq:O8} without prefactors.

We define the dimensionless functions $G_\iota$, with $\iota \in \{ 1,2,3,T\}$, as follows\footnote{The factor $c_V$ is inserted to absorb trivial factors
due to the $\omega \sim (\bar u u + \bar d d)/\sqrt{2}$ and  $\rho^0 \sim (\bar u u - \bar d d)/\sqrt{2}  $ wave functions. $c_V= - \sqrt{2}$ for $\rho$ in $b \to d$ transitions, $c_V = \sqrt{2}$ in 
all other transitions into $\omega$ \& $\rho^0$ and $c_V=1$ otherwise. }:
\begin{eqnarray}
\label{eq:FFdec}
c_V \, {\cal A}^{*\rho}(V)  &=&
 k_G \,  \Big(G_1(q^2) P_1^\rho +
 G_2(q^2) P_2^\rho + G_3(q^2) P_3^\rho   \Big) \; \nonumber  \\[0.2cm]
 {\cal A}^{*\rho}(P)   &=&
k_G \,  \Big( G_T (q^2) P_T^\rho \Big) \;,
 \end{eqnarray}
 with $k_G = -2 e/g$ to be explained further below.
  The transverse ($ q_\rho P_\iota^\rho = 0$) Lorentz structures $P_{i,T}$, of mass dimension $[P_i] = 2$ and $[P_T]=1$, 
 are given in appendix \ref{app:projectors}. 
The physical domain of the  $B \to P(V) \gamma^* \to P(V) l^+,l^-$-transition 
 is   $(2m_l)^2 \leq q^2 \leq (m_B - m_{P,V})^2$, with $l$ being a lepton\footnote{Analytic continuation to other values of $q^2$ can be related to other processes, e.g. $B + V \to \gamma^*$
by crossing symmetry. 
The domain of validity of our computation is discussed in section \ref{sec:numerics}.}.
Under exchange of chirality  $(1+\gamma_5) \to (1-\gamma_5)$ in ${\cal O}_8$ \eqref{eq:O8}, often denoted as ${\cal O}'_8$, the $G_\iota$-functions transform as follows:
\begin{equation}
\label{eq:prime}
\{G_1,G_2,G_2,G_T\}  \quad \stackrel{(1+\gamma_5) \to (1-\gamma_5)}{ \to} \quad  \{G_1,-G_2,-G_3,G_T\} \;,
\end{equation}
at leading order in the weak interactions.
Thus $G_1$ and  $G_T$ are parity conserving and $G_2$ and $G_3$ are parity violating.
 The operator $\mathcal O_8$ \eqref{eq:O8} 
is consistent with the effective Hamiltonian 
\begin{align}
\mathcal H_{\mathrm{eff}} &= -\frac{G_F}{\sqrt 2}V_{ts}^* V_{tb} ( C_7 \mathcal O_7 + C_8 \mathcal O_8)+ ...\,, &
\mathcal O_7 &= -\frac{e m_b}{8\pi^2}\bar s \sigma\cdot F (1+\gamma_5)b \;.
\end{align}
In the case of $D\rightarrow M\gamma^*$ the replacements $b\rightarrow c$, $m_b\rightarrow m_c$ and
$V_{ts}^* V_{tb}\rightarrow V_{cb}^* V_{ub}$ are used.
The normalisation constant
\begin{equation}
k_G \equiv -2 \frac{e}{g}
\end{equation}
used in \eqref{eq:FFdec} is chosen such that
$G_\iota$-functions  parallel the  
 standard vector $T_i$ and pseudoscalar $f_T$ penguin form factors in the amplitude:
\begin{align}
\matel{\gamma^*(q,\rho) V(p,\eta)}{H_{\rm eff}}{\bar B} &\propto \sum_i (C_7 T_i(q^2) + C_8 G_i(q^2)) P_i^\rho + \dots \nonumber \\
\matel{\gamma^*(q,\rho) P(p)}{H_{\rm eff}}{\bar B} &\propto (C_7 f_T(q^2) + C_8 G_T(q^2)) P_T^\rho + \dots
\label{eq:analogue}
\end{align}

\subsection{The sum rule}

The matrix elements \eqref{eq:schematic} are extracted from the following correlation function\footnote{For the sake of notational simplicity we shall keep the photon polarisation tensor contracted here as with respect to \eqref{eq:matel}, though from a physical point of view this
does not make sense  for an off-shell photon.}:
\begin{equation}
\label{eq:CF}
\Pi^V(q^2,p_B^2) =  \epsilon^{*\rho}(q) \Pi_\rho^V(q^2,p_B^2) =   i \int_x \matel{\gamma^*(q) V(p) }{T J_B(x) \tilde O_8(0)}{0} e^{-i p_B \cdot x} \;,
\end{equation}
where the $B$-mesons figures as an interpolating current:  
\begin{equation}
\label{eq:JB}
J_B  = i m_b \bar b \gamma_5 q \;, \qquad \matel{\bar B(p_B)}{J_B}{0} = m_B^2 f_B  \;.
\end{equation}
In the equation above $q = u,d$ are light flavoured quarks and $f_B$ is the standard $B$-meson decay 
constant.

\begin{figure}[h]
 \centerline{\includegraphics[width=5.5in]{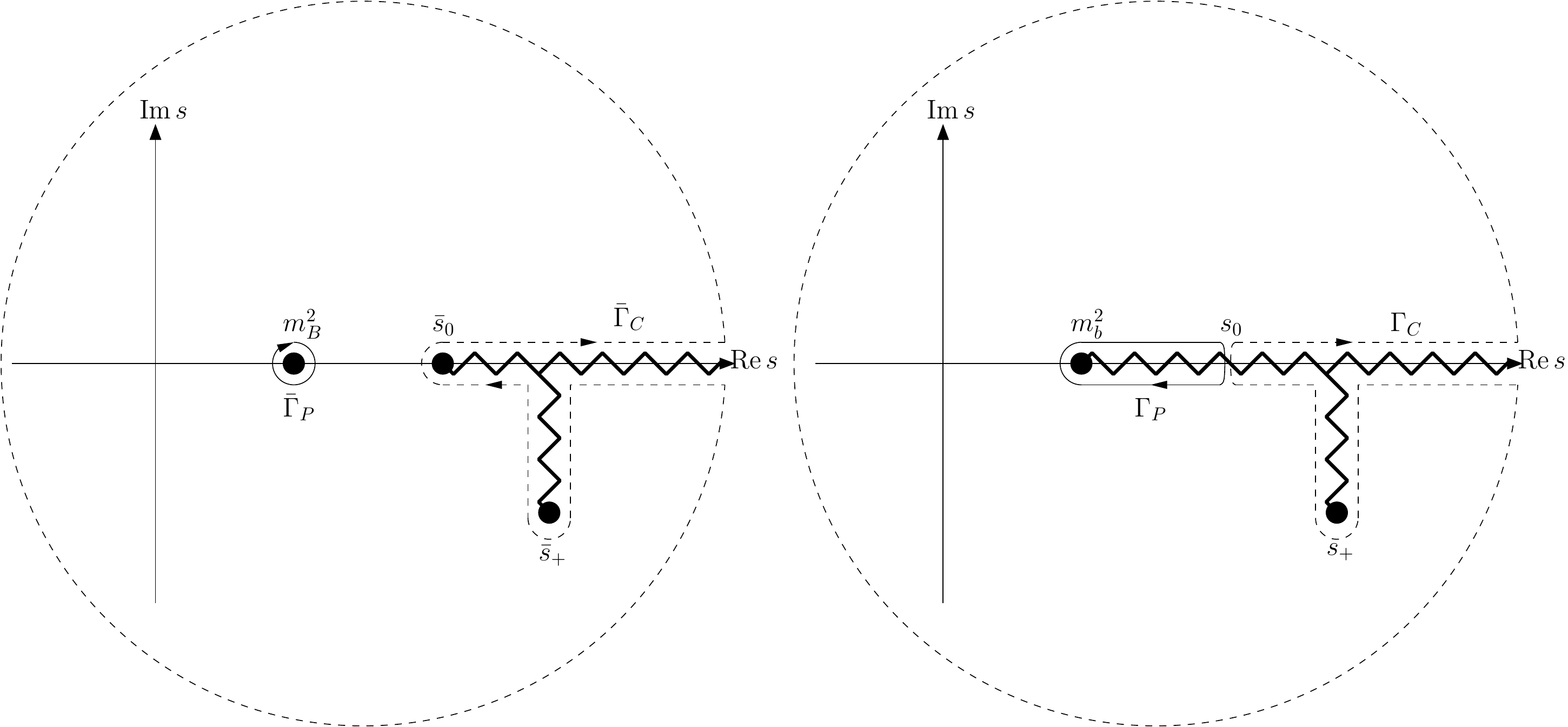}
}
 \caption{\small $\Gamma_{\rm P}[\overline \Gamma_{\rm P}]$ and $\Gamma_{\rm C}[\overline \Gamma_{\rm C}]$ correspond to the straight and dashed paths  in the right[left] figure respectively.
 (left) Analytic structure of the correlation function in QCD.  There is an isolated $B$-pole at $s = m_B^2$ 
 and a  branch point $\bar s_0 = (m_B+2m_\pi)^2$ at  the continuum threshold.  Furthermore the existence of  a complex branch point $\bar s_+$, which corresponds to an anomalous threshold, 
 can be inferred from the work of K\"all\'{e}n \&  Wightman c.f. appendix \ref{app:KW}.
 The path $\overline \Gamma = \overline \Gamma_P \cup \overline \Gamma_{\rm C}$ is a possible path for Eq.~\eqref{eq:SR1}. 
 (right) Analytic structure of the correlation function as found in leading order perturbation theory. The branch point related to the normal threshold starts at $m_b^2$. The existence of the anomalous branch point $s_+$ is shown in appendices \ref{app:Leq} and \ref{app:explCS} respectively. 
The two branch points $\bar s_+$ and $s_+$ are expected to be close, but not identical, 
in the same as $m_B^2$ is close to $m_b^2$.}
 \label{fig:contours}
 \end{figure}

Leaving aside 
the issue of parasitic cuts and how to compute the correlation function to the next section, we may apply 
 standard techniques of dispersion relations and Borel transformations \cite{SVZ1}
to extract the matrix element under consideration.
The  dispersion representation of the correlation function in the variable $p_B^2$\footnote{Possible subtraction terms, due to ultraviolet (UV)-divergences, are ignored in view of the fact that they disappear 
under Borel-transformation.},
\begin{equation}
\label{eq:SR1}
\Pi^V(q^2, p_B^2) =  \frac{1}{2 \pi i} \oint_{\overline \Gamma} \frac{ds \Pi^V(q^2, s) }{s-p_B^2}   \;,
\end{equation}
is nothing but  Cauchy's integral theorem: The closed path $\overline \Gamma$ is chosen such that no 
singularities, including anomalous thresholds (to be discussed in the next section), are crossed. 
An example is shown in Fig.~\ref{fig:contours} for the analytic structure of the correlation function in 
QCD; $\overline \Gamma = \overline \Gamma_{\rm P} \cup \overline \Gamma_{\rm C}$.
In a second step  advantage is taken of the isolated $B$-pole 
by splitting the dispersion integral into two parts as follows,
\begin{equation}
\label{eq:SR1b}
\Pi^V(q^2, p_B^2) = \frac{m_B^2 f_B}{m_B^2 - p_B^2}  \matel{\gamma^*(q) V(p)}{\tilde O_8}{\bar B(p_B)} 
+   \frac{1}{2 \pi i}   \oint_{\overline \Gamma_{\rm C}}  \frac{ds \Pi^V(q^2, s) }{s-p_B^2} \;.
\end{equation}
Equating \eqref{eq:SR1} and \eqref{eq:SR1b} one obtains:
\begin{equation}
\label{eq:SR1c}
\frac{m_B^2 f_B}{m_B^2 - p_B^2}  \matel{\gamma^*(q) V(p)}{\tilde O_8}{\bar B(p_B)}  = 
  \frac{1}{2 \pi i} \left(   \oint_{\overline \Gamma}  \frac{ds \Pi^V(q^2, s) }{s-p_B^2}  -    \oint_{\overline \Gamma_{\rm C}}  \frac{ds \Pi^V(q^2, s) }{s-p_B^2} \right) \;.
\end{equation}
For the purpose of numerical improvement a Borel transformation,  \begin{equation}
\label{eq:Borel}
B_{s \to M^2}[ \frac{1}{x-s}]  = \frac{e^{-x/M^2}}{M^2} \;,
\end{equation}
in the variable $p_B^2$ applied 
to \eqref{eq:SR1c} to obtain:
\begin{equation}
\label{eq:exact}
 \matel{\gamma^*(q) V(p)}{\tilde O_8}{\bar B(p_B)}  =  
D[\Pi^V,\overline \Gamma] - D[\Pi^V,\overline \Gamma_{\rm C}] \;,
\end{equation}
where we introduce the shorthand notation:
\begin{equation}
D[f,\Gamma_f] \equiv  \frac{1}{f_B m_B^2} \frac{1}{2 \pi i}  \oint_{\Gamma_f  } 
 ds e^{(m_B^2-s)/M^2}f(q^2, s)  \;.
\end{equation}
The expression in \eqref{eq:exact}, up to neglecting the width of the $B$-meson, is exact although rather cryptic.
Approximations enter the calculation of the correlation function $\Pi^V$ due to neglecting 
higher twist and $\alpha_s$-corrections  and in estimating   $D[\Pi^V, \overline \Gamma_{\rm C}]$. 
Let us be more precise about the latter point. 
Whereas $D[\Pi^V,\Gamma] \approx D[\Pi^V|_{\rm LC-OPE},\overline \Gamma]$ is a good approximation for off-shell $p_B^2$ up to the
truncations in twist and $\alpha_s$ mentioned above, the approximation 
$D[\Pi^V,\overline \Gamma_{\rm C}] \approx D[\Pi^V|_{\rm LC-OPE},\Gamma_{\rm C}]$, which goes under the name of \emph{semi-global quark hadron duality}, is less transparent and usually the main limitation of a sum rule computation.   In the full theory $\overline \Gamma_{\rm C}$
marks the onset of the continuum threshold which corresponds to 
the lowest lying multiparticle state (e.g. $\bar s_0=(m_B+2m_\pi)^2$ in QCD\footnote{In principle there might be further isolated states, with $B(0^-)$ quantum numbers, between $m_B^2$ and $(m_B+ 2 m_\pi)^2$.
Note there are non listed in PDG \cite{PDG}.  
In our discussion those states would simply be included into  the path $\bar \Gamma_C$.}). 
For the LC-OPE dispersion representation one introduces an effective continuum   threshold $s_0$ \cite{SVZ1,CK00}\footnote{Whereas  $s_0 \approx \bar s_0$  ought to be the case 
exactness cannot be expected to hold. 
Realistically one can expect  
$s_0$ to be somewhere between say $(m_B+2 m_\pi)^2$ and $(m_B + m_\rho)^2$.
Whether or not this affects the final result depends on the convergence of the LC-OPE and Borel parameters
and has to be analysed and is discussed in section \ref{sec:numerics}}, which corresponds to the duality approximation mentioned above. 

The crucial point in connection with the anomalous threshold,
which results in branch cuts extending into the complex 
plane, is that its real part is above the continuum threshold, $m_b^2 + m_B^2/2 > s_0$,
and therefore it is entirely included in $\Gamma_{\rm C}$ and will not contribute to the final sum rule\footnote{
It is also suppressed by the Borel transformation, both due to the large real part of $s$ and the oscillation in the exponential due to $\Im\,s\neq 0$ along
the associated branch cut.}.
Therefore the path $\Gamma$ minus the path $\Gamma_{\rm C}$ corresponds to 
the path $\Gamma_{\rm P}$ that encircles the real line segment from  $m_b^2$ to $s_0$. The final sum rule
can be written as   
\begin{alignat}{2}
&  \matel{\gamma^*(q) V(p)}{\tilde O_8}{\bar B(p_B)} \; &\simeq&\;  D[\Pi^V|_{\rm LC-OPE},\Gamma] -   D[\Pi^V_{\rm LC-OPE},\Gamma_{\rm C}]  \nonumber \\[0.1cm]
& \quad  = D[\Pi^V|_{\rm LC-OPE}, \Gamma_P ]  \; &=& \;   \frac{1}{f_B m_B^2} \int_{m_b^2}^{s_0} ds e^{(m_B^2-s)/M^2} \rho^V(q^2,s)
\label{eq:SR2}
\end{alignat}
with 
\begin{equation}
\label{eq:disc1}
2 \pi i  \rho^V(q^2,s) =   \text{Disc}_s\Pi^V(q^2,s)  = 
  \Pi^V(q^2,s+i0) -  \Pi^V(q^2,s-i0) \;, 
\end{equation}
where we have dropped the subscript LC-OPE in \eqref{eq:disc1}.  
Note the radius of the path $\Gamma_{C}$ and $\Gamma$ (as well as for the barred quantities) 
does not enter the final relation \eqref{eq:SR2}. 
The important point that the endpoint of the duality interval is much larger than the intrinsic scale of QCD; $s_0  \gg \Lambda_{\rm QCD}^2$.

\subsubsection{Remarks on dispersion relations and anomalous thresholds}
\label{sec:disp_anomal}

As the appearance of complex singularities in forms of anomalous thresholds 
is rather non-standard in sum rule computations, we consider it worthwhile 
to add a few remarks.  Our three main points are:
\begin{itemize}
\item  We note, again, that Eq.~\eqref{eq:SR1} is nothing but the application of 
Cauchy's integral theorem. Thus  knowledge of the analytic structure 
of the correlation function is mandatory.  
\item 
The existence of the pole at the $B$-meson mass\footnote{Ignoring the finite width, which otherwise move the pole into the lower half plane of the second Riemann-sheet.}
and its residue in terms of the matrix elements in Eq~\eqref{eq:SR1b} 
can  be inferred from derivations 
like the one presented in chapter 10.2 in \cite{Weinberg1}. 
\item The part not related to the $B$-meson pole, i.e. the part encircled by 
$\Gamma_{\rm C}$, is to the RHS of ${\rm Re}[s_0]$.
 In practice this means that it is suppressed,  
by the Borel transformation \eqref{eq:Borel},  by at least a factor of $e^{(m_B^2-s_0)/M^2}$ with respect 
to the $B$-pole part.
\end{itemize}

A few remarks on the connection between physical states and singularities: 
For a two-point function,\footnote{In this paragraph it is assumed that the operators
are gauge invariant.}  a dispersion representation 
is in one-to-one correspondence with the insertion of a complete set of states 
as is explicit in the celebrated K\"all\'{e}n-Lehmann representation \cite{KL} and derivations 
thereof.
Thus, the analytic structure, in the complex plane of the four momentum invariant, 
has a cut and poles on the real line starting from the lowest state in the spectrum.
For correlation functions with three and more fields, there is no such direct relation. 
The analytic structure can be more involved as singularities other than those related to intermediate 
states might appear, known as anomalous thresholds e.g. \cite{Smatrix,Todorov}.
Singularities related to unitarity, that is to say to an insertion of a complete set of 
states, are called normal thresholds.
From the viewpoint of a dispersion relation, normal and anomalous thresholds should
 be viewed as being on the same footing\footnote{Let us add that even among the normal thresholds there are states 
which do not correspond to the insertion of a single identity. E.g. the parasitic states which 
correspond to different time ordering. As discussed in this paper they do appear when 
no momentum is flowing into one of the operators of the correlation function.}
as only the analytic structure counts.  
Which singularities are relevant for the physics in question is another matter. 
Clearly, here  we are interested in the matrix element corresponding to the residue of
the pole of the $B$-meson which belongs to the normal part. The arguments 
above should make it clear that the anomalous thresholds do no more harm than
any other continuum contribution to the extraction of the matrix element in question.

\section{The computation}
\label{sec:computation}

\begin{figure}[h]
 \centerline{\includegraphics[width=6.0in]{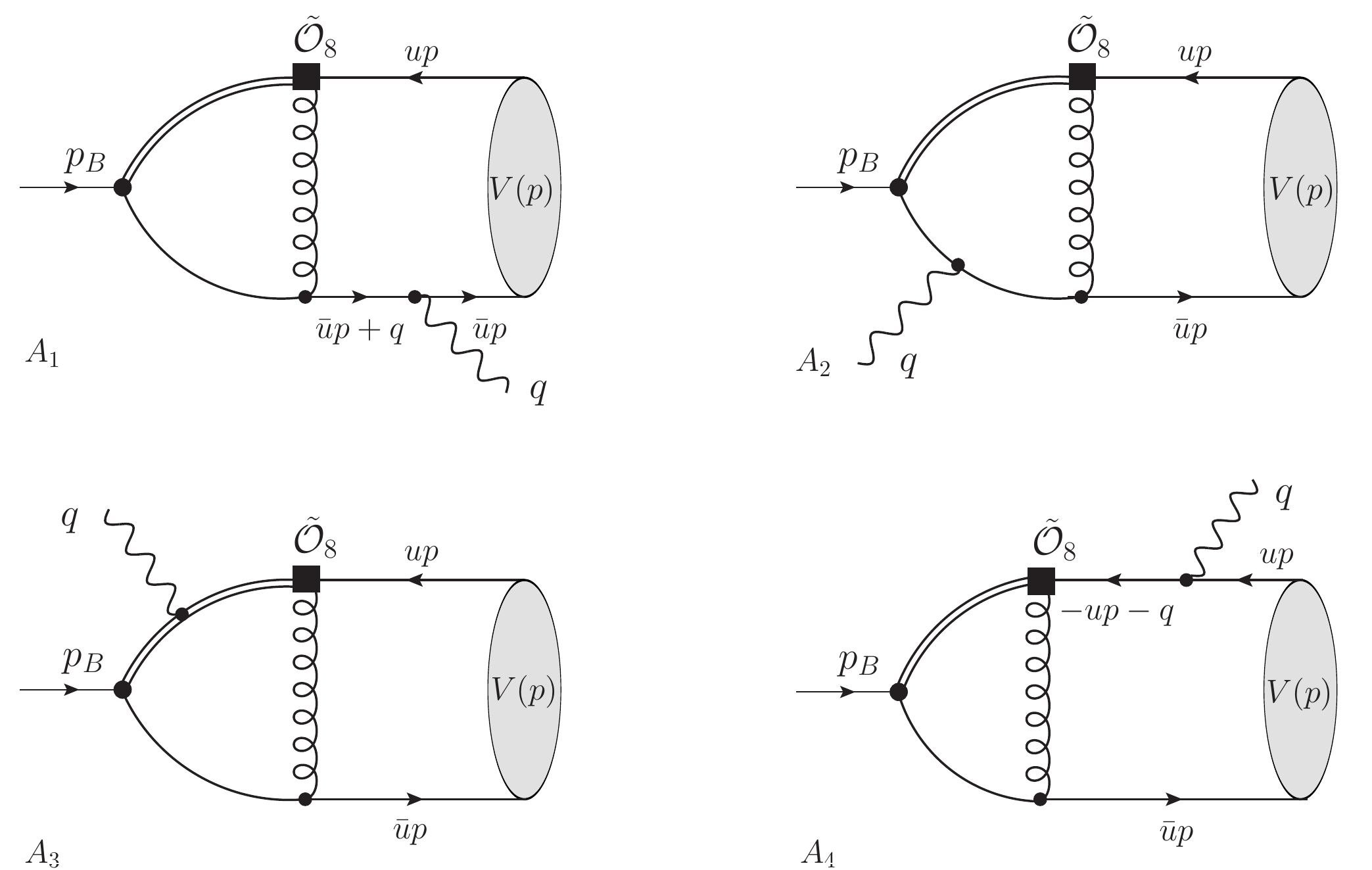}}
  \centerline{
 \includegraphics[width=6.0in]{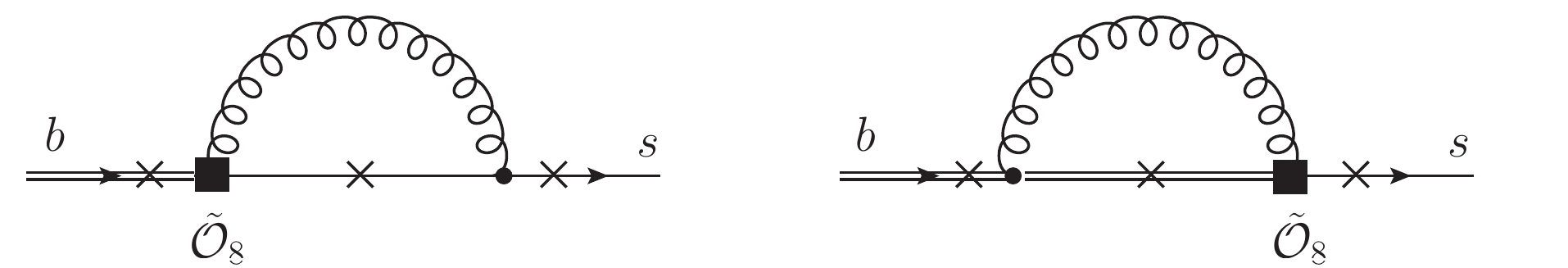}}
 \caption{\small (top/middle) Diagrams $A_1$ to $A_4$,
 correspond to all four possibilities with the gluon from 
 the weak vertex connecting to the spectator quark. (bottom) Non-spectator corrections. They have been computed in \cite{AAGW01} and factorise into a form factor and $B \to V/P$-form factor as described in appendix \ref{app:non-spectator}.
The crosses indicate all possible photon insertions.} 
 \label{fig:diaA}
 \end{figure}

In this section we provide some more details of the computation with some explicit results deferred to the appendices.
At leading order in $\alpha_s$ there are a total of twelve graphs. 
They can be split  into those where the gluon connects to the spectator (s)  and the ones where it connects to 
the non-spectator (ns) quark:
\begin{equation}
\label{eq:decsns}
G_\iota(q^2)  = G_\iota^{(s)}(q^2) + G_\iota^{(ns)}(q^2) \;.
\end{equation}
The four diagrams denoted by  $A_1$ to $A_4$  in Fig.\ref{fig:diaA}(top,middle) contribute to 
$G_\iota^{(s)}$ whereas the diagrams at the bottom of the same figure correspond to the 
$G_\iota^{(ns)}$-contributions. Hereafter we use $\bar u \equiv 1 -u$.   The $G_\iota^{(ns)}$-functions  factorise into a function $f(q^2/m_b^2)$ times
the standard vector, axial or tensor form factors. The function $f$ has been obtained in  
the inclusive case in  \cite{AAGW01}\footnote{We would like to add that it would be possible 
to  compute these contribution within LCSR itself.}, in terms of an expansion in powers of 
$q^2/m_b^2$ and logarithmic terms. 
The two diagrams where the gluon connects to the non-spectator quark  and 
 photon emission from the latter  are not shown.  These diagrams are expected to be small,
since no fraction of the $m_b$-rest mass is transmitted to the energetic photon and we shall 
neglect them. For the same reason and for being of higher twist we expect the diagrams where the  gluon is radiated into the final state meson to be suppressed\footnote{A rough estimate can be given by comparing the similar case where
a gluon is radiated from a charm loop, instead of ${\cal O}_8$ to the hard spectator or the final state meson.
Taking the estimates of \cite{BB01} and \cite{BZtime,BJZ} we find roughly a factor of four between them.}.

\subsection{The problem of parasitic cuts}
\label{sec:parasite}

Due to the fact that there is no momentum flowing into the weak vertex 
at $\tilde {\cal O}_8$, there's an ambiguity in  separating the cuts corresponding 
to the $B$-meson from other cuts.  The general problem originates from the fact that the relation 
between correlation functions of higher degree and  matrix elements is complicated by time ordering and 
a non-trivial analytic structure. Similar issues appear in euclidean field theory and represent an obstacle  to
extract  matrix of more than two hadronic states from correlation functions on the lattice\cite{MT90}.
In LCSR the problem is best understood by first introducing its (partial) cure.

\begin{figure}[h]
 \centerline{\includegraphics[width=4.2in]{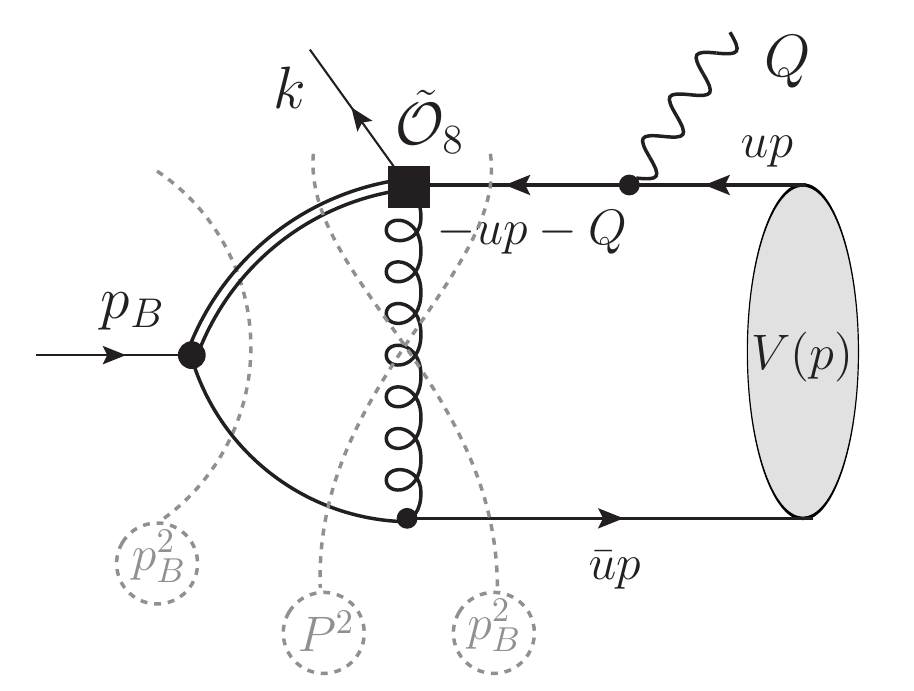},}
 \caption{\small Various cuts in the variables $p_B^2$ and $P^2 \equiv (p_B-k)^2$. The cut in $P^2$ is of a parasitic type in the sense that for $k \to 0$ it cannot be distinguished from $p_B^2$ yet it clearly not associated with the $B$-meson as it does not cut in the $b$-quark line. 
 The two cuts in $p_B^2$ are of the $2$-parton and $3$-parton type and should and are both included. Here and thereafter the double-line denotes the $b$-meson propagator.}
 \label{fig:parasite}
 \end{figure}

We follow the method introduced by Khodjamirian for 
$B \to \pi \pi$ \cite{K00}, which might be seen as an extension of earlier ideas  \cite{russians}, 
and introduce a spurious momentum $k$
into the weak vertex. 
This introduces two further  momenta denoted by $P = p_B - k$ and $Q = q - k$.
Formally, the $1 \to 2$ decay is augmented by the spurious momentum $k$  to a $2 \to 2$ scattering process which has six
independent kinematic variables: 
$\{ q^2, Q^2, p_B^2, P^2 , k^2, p^2 \}$.  
Without any consequence for our purposes we can set $q^2 = Q^2$ and 
$k^2 = 0$.  Capital $Q$ will from now on only be used for the four momentum throughout the  paper.
Recalling that  $p^2 = m_{P,V}^2$  the six kinematical invariants are reduced 
to $\{q^2,P^2,p_B^2\}$ which we shall discuss in the next section.
The variable  $P^2$ remains the only trace of the spurious momentum at this stage.
How it effectively disappears from the final result is discussed in 
the next subsection after we discuss the light-like dominance of the correlation function.
At the level of the correlation function \eqref{eq:CF}
the change is implemented by changing the photon momentum $q \to Q$.
The above mentioned cuts then branch into cuts in $p_B^2$ and $P^2$, c.f. Fig.~\ref{fig:parasite} ,where the former correspond to the $B$-meson and the latter to parasitic ones.  

The extension of the Lorentz-structures to the case where we include the spurious momentum $k$ is given in appendix \ref{app:projk}. 
Using the latter  we parametrise the correlation functions as follows:
\begin{eqnarray}
\label{eq:CFwith_p}
\Pi^V =  \sum_{i=0}^4  g_i (q^2) \epsilon(Q) \cdot p_i  \;, \qquad 
\Pi^P =  \sum_{i \in \{0,T,\bar T\}}  g_i (q^2) \epsilon(Q) \cdot p_i \;,
\end{eqnarray}
where $\epsilon(Q)$ is the photon polarization tensor and 
$(p_0)_\rho  = Q_\rho$ is a non-transverse structure related to contact terms.
As previously stated  the Lorentz structures corresponding to the $G_\iota$-functions are transverse 
even for an off-shell photon. 
This is  not necessarily true for  the correlation function. 
Why these terms are there and why they do not affect the extraction of the matrix element 
is discussed in appendix \ref{app:WTI} in terms of a Ward  Takahashi identity (WTI).

\subsection{The light-cone expansion}

The correlation function is expected to be dominated by light-like distances in the case where
the kinematical invariants $k^2$, $q^2$, $p_B^2$ and $P^2$ \footnote{The remaining two invariants are
$Q^2 = q^2$ and $p^2 = m_{P,V}^2$. The former thus does not necessitate a separate statement and the latter is on-shell by virtue of being the momentum of a physical state.}
are below the thresholds. 
In that case, the  light-cone operator product expansion (LC-OPE),
c.f. \cite{CK00} for a review article on the topic, is applicable. 
For the physical matrix element $q^2$ and $P^2$ necessitate analytic continuation, an issue 
which we defer to  sections \ref{sec:strong} and \ref{sec:q2-range} respectively.
Schematically the LC-OPE reads as follows:
\begin{equation}
\label{eq:LC-OPE}
\Pi(q^2,p_B^2,P^2) =  \sum_i T^{(i)}_H(q^2,p_B^2,P^2;\mu_F;u) \circ \phi^{(i)}(u,\mu_F) \;,
\end{equation}
where $i$ sums over different distribution amplitudes (DAs) of increasing twist. The twist corresponds to 
the dimension of the operator minus its spin. The terms are suppressed by $\Lambda_{\rm QCD}$ over the virtuality to the power of the twist.  In this work we limit ourselves to the the leading twist-$2$. The 
relevant DAs are summarised in appendix \ref{app:DA}.
The variable $u$ represents generic parton momentum fractions,
 the symbol $\circ$ stands for the 
integration over the latter and 
$T_H$ is a perturbatively calculable hard kernel. The symbol $\mu_F$ denotes the collinear factorisation 
scale and separates, within 
the LC-OPE, the SD physics in the kernel $T_H$ from the LD part in the DA.  This scale should not be confused with the renormalisation scale $\mu_{\rm UV}$ to be discussed in the numerics section.
For the computation we use FeynCalc \cite{Mertig:1990an}.
We would like to highlight two issues in connection with the calculation:
\begin{itemize}
\item \emph{Infrared (IR)-divergences:}
We note that the diagram $A_2$ in Fig.~\ref{fig:diaA} has a potential soft divergence for $p^2 \to 0$ 
and a collinear divergence for $q^2 \to 0$. 
The former cancels and the latter appears only in the $P_3$ and $P_T$ Lorentz-structures 
which do not contribute at $q^2 = 0$.
\item 
\emph{Schouten identity:}
For structures like $Q_\rho \epsilon(\eta,p,p_B,Q)$ the Schouten identity 
$g^{ab} \epsilon^{cdef} 
 = g^{ac} \epsilon^{bdef} - g^{ad} \epsilon^{bcef} -  g^{ae} \epsilon^{bdcf} - g^{af} \epsilon^{bdef}$ has to be used since they contain
pieces of the Lorentz-structure $(p_1)_\rho$ in \eqref{eq:Vprojk}.
\end{itemize}
UV-divergences are present in diagrams $A_2$ and $A_3$ but are of no consequence 
as the discontinuities of the correlation functions do not depend on them.
Explicit results in terms of Passarino-Veltman (PV) functions \cite{Passarino:1978jh} and their corresponding dispersion relations, including the handling of the
 complex branch cuts, are given appendices 
\ref{app:results} and \ref{app:dispersion} respectively.

\subsection{Analytic continuation and appearance of strong phases}
\label{sec:strong}

As previously stated the LC-OPE is valid when all invariants take on values such that no thresholds are crossed. 
To obtain a physical result two of those invariants, $q^2$ and $P^2$, need to be analytically continued:
$q^2$ to enter the physical domain for $B \to V(P) ll$ transitions and $P^2$ to eliminate the 
spurious momentum $k$.

\begin{figure}[h]
 \centerline{\includegraphics[width=6.6in]{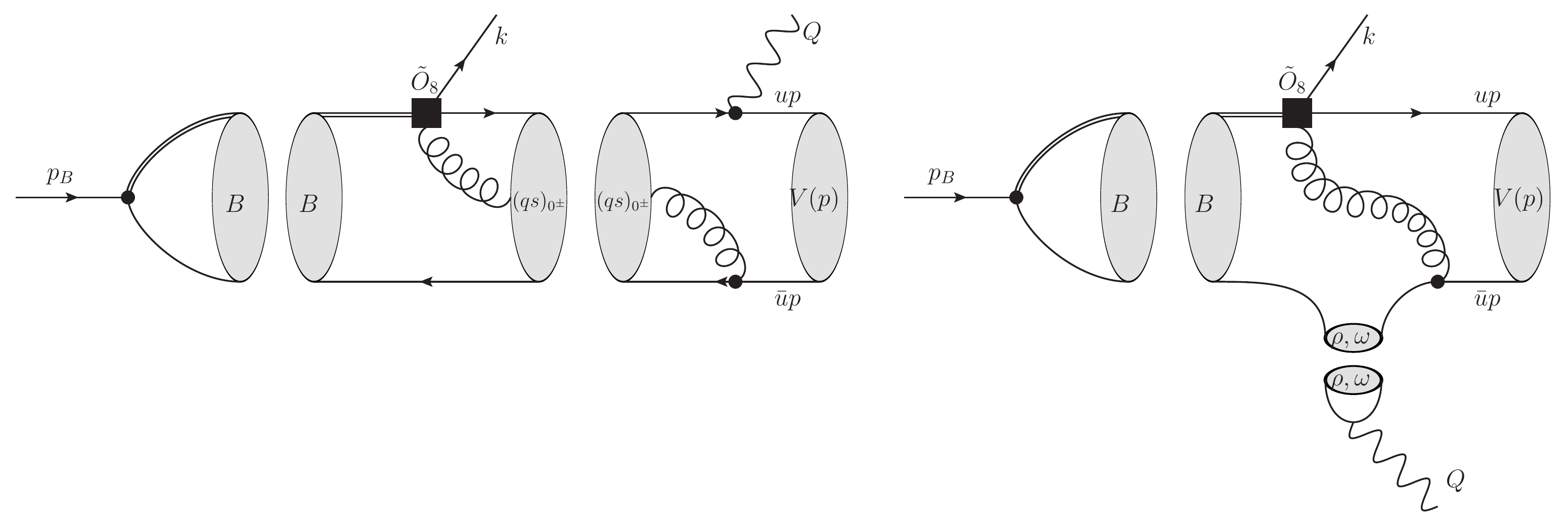}}
 \caption{\small (left) Hadronic interpretation of the $3$-particle cut 
 in Fig.\ref{fig:parasite}  in terms of a LD hadronic process. The latter is a source 
 for the strong (CP-even) phase that we obtain for the  $G_\iota(q^2)$-functions. 
 (right) Hadronic interpretation of the strong phase due to $q^2 > 0$, 
associated with $B \to V (\rho,\omega)  \to V \gamma^*  \to V ll$-type transitions.}
 \label{fig:LD}
 \end{figure}

For $B \to V(P) ll$ the physical range for $q^2$ is between $ (2m_l)^2$ and $(m_B - m_{P,V})^2$ and it has become 
customary to exclude the region below $1 \GeV^2$ in order to avoid the  
$(\rho,\omega)$-resonance region.  For $B \to V \gamma$, which corresponds to $q^2 = 0$, it can be
argued that one is still considerably low\footnote{$q^2 = 0$ is sufficiently  below the  $(\rho,\omega)$-threshold  region and therefore the LC-OPE is expected to work.}. Some more details, concerning individual 
graphs and the high $q^2$ region can be found in section \ref{sec:q2-range}.
As previously stated, the only trace of the spurious momentum 
is in $P^2 \equiv (p_B-k)^2 \neq p_B^2$. 
This trace can be lifted by analytically continuing $P^2 \to m_B^2 + i0$. 
Note that if we had the full solution of the correlation function, then 
$p_B^2 = m_B^2$  would lead 
to an exact projection by virtue of an LSZ-reduction.  
In the sum rule approximation the remnant of this is   
the fact that the integral representation \eqref{eq:SR2} averages over a narrow range of  $m_B^2$.
On the level of the LC-OPE, this analytic continuation is expected to hold as it is far above all thresholds; the 
variable $P^2$ does not cut through the $b$ quark line c.f. Fig.~\ref{fig:parasite}. 
Both analytic continuations lead to LD contributions which in turn lead to strong phases.
This is illustrated for a $P^2 = m_B^2$-cut in Fig.~\ref{fig:LD}(left) and for a $q^2 \simeq m_\rho^2$ cut in
Fig.~\ref{fig:LD}(right). 

In summary, both $q^2$ and $P^2$ are analytically continued sufficiently far above the thresholds, 
much alike the open charm region in $e^+e^- \to  (\bar c c) \to e^+e^-$. 

\section{Results, summary and numerics}
\label{sec:numerics}

We note that  in the sum rule the product $[m_B^2 f_B]  \times 
\matel{\gamma^*(q) V(p)}{\tilde O_8}{\bar B(p_B)}  $, c.f. Eq.~\eqref{eq:SR1b}, 
rather than the $G_i(q^2)$ functions themselves are extracted. This  suggests that one should use a sum rule
determination of the same order in the quantity $[m_B^2 f_B]\;$\footnote{This quantity corresponds to the
matrix element of the interpolating current \eqref{eq:JB}.} in order
to extract the matrix element(s). Such a strategy has for example been 
proposed in \cite{BBB97}.
From Fig.~\ref{fig:parasite} it is evident that the $2$-particle cut 
corresponds to a decay constant of order ${\cal O}(\alpha_s^0)$. 
The $3$-particle cut in the same figure corresponds partially to an ${\cal O}(\alpha_s)$-correction. We expect the former to be dominant so we feel justified to use the sum rule result to order ${\cal O}(\alpha_s^0)$,

\begin{eqnarray}
\label{eq:fBSR0}
[m_{B_q}^2 f_{B_q}]^2|_{SR_0} &=& (m_b+m_q)^2 e^{\frac{m_B^2-m_b^2}{M^2}} 
\Big(  - m_b\aver{\bar qq}_\mu 
- \frac{m_b}{2 M^2} (1 - \frac{m_b^2}{2 M^2}) \aver{ \bar q G q}_\mu 
\\[0.1cm]
   &+&   \frac{3}{8 \pi^2} 
\int_{(m_b+m_q)^2}^{s_0}  \!\!\!\!\!\!\!\!\!\!\!\! e^{\frac{m_b^2-s}{M^2}  }   
(s\!-\!(m_b\!-\!m_q)^2) \sqrt{(s\!-\!m_b^2\!-\!m_q^2)^2\! -\! 4 m_b^2 m_q^2}  \frac{ds}{s}\Big)
\nonumber 
\end{eqnarray}
which has 
been known for a long time \cite{AE83}. The parameters $M^2 = M^2[f_{H_q}]$ and $s_0 = s_0[f_{H_q}]$ 
are not, necessarily, the same as the ones in the sum rule for $G_\iota$-functions. Further discussion is 
deferred to appendix \ref{app:input}.

Following the decomposition \eqref{eq:decsns} at twist-2 the spectator parts decompose for the vector and pseudoscalar final state as follows:
\begin{eqnarray}
\label{eq:Gdec}
G_i^{(s)}  &=&   G_i^{(\perp)}(q^2) + G_i^{(\parallel)}(q^2)   \;,   \nonumber \\[0.1cm]
G_T^{(s)} &=& G_T^{(P)}(q^2)    \;,
\end{eqnarray}

\begin{table}[tb]
\addtolength{\arraycolsep}{3pt}
\renewcommand{\arraystretch}{1.2}
$$
\begin{array}{ l | c |  cccccccc }
 &  & \rho[\pi]^+ &\rho[\pi]^0,\omega &\rho[\pi]^- & K^{*}[K]^{+} &  K^{*}[K]^{0} &  K^{*}[K]^{-} &
 \bar K^{*}[\bar K]^{0} & \phi\\ \hline
 &  & (u \bar d) & (\bar u u ) \pm (\bar d d ) & (u \bar d) & (u \bar s) & (d \bar s) & (s \bar u) & (s\bar d) & (s \bar s) \\ \hline
B^-  & (b \bar u ) &  - & - & b\to d & - & - & b\to s & - & -\\
\bar{B}^0 & ( b \bar d)  & -  & b\to d & - & - & - & - & b\to s & -\\
\bar{B}_s & (b\bar s)  & - & - & - & -  & b\to d & - & - & b\to s\\ 
D^0 & (c \bar u) & - & c \to u & - & - & - & - & - & - \\
D^+ & (c \bar d) & c \to u & - & - & - & - & - & -& - \\
D_s & ( c \bar s) & - & - & - & c \to u & - & - & - & - \\
\end{array}
$$
\addtolength{\arraycolsep}{-3pt}
\renewcommand{\arraystretch}{1}
\caption[]{\small FCNC-transitions up to charge conjugation for $B(D) \to V(P)$ as indicated. The valence quark content of the mesons are indicated in brackets
and the type of transition is indicated. 
We do not consider $\eta$ and $\eta'$ for the pseudoscalars. There are a total of $11_V + 8_P = 19$-transitions.}\label{tab:0}
\end{table}

\begin{table}
\addtolength{\arraycolsep}{3pt}
\renewcommand{\arraystretch}{1.3}
$$
\begin{array}{lccl|lccl}
 & G_1^{(\perp)}(0) \cdot 10^2 & \text{unc.}\% & \text{type} & & G_1^{(\perp)}(0) \cdot 10^2 & \text{unc.}\% & \text{type}
\\ \hline

 B^-\rightarrow\rho^-\gamma & 0.29-0.39i & 25\% & (bD)^-  & 
 \bar B_s\rightarrow K^{*0}\gamma & 0.21+0.18i & 27\% & (bD)^0  \\
 B^-\rightarrow K^{*-}\gamma & 0.29-0.40i & 26\% & (bD)^-  & 
 \bar B_s\rightarrow\phi\gamma & 0.26+0.23i & 26\% & (bD)^0  \\
 \bar B^0\rightarrow\rho^0\gamma & 0.22+0.19i & 27\% & (bD)^0  & 
 D^0\rightarrow\rho^0\gamma & -7.0-5.0i & 32\% & (cu)^0  \\
 \bar B^0\rightarrow\omega\gamma & 0.19+0.17i & 33\% & (bD)^0  & 
 D^0\rightarrow\omega\gamma & -6.1-4.3i & 34\% & (cu)^0  \\
 \bar B^0\rightarrow\bar K^{*0}\gamma & 0.20+0.20i & 28\% & (bD)^0  & 
 D^+\rightarrow\rho^+\gamma & -1.9+2.5i & 32\% & (cu)^+  \\&&&& 
 D^+_s\rightarrow K^{*+}\gamma & -1.8+2.1i & 33\% & (cu)^+ 
\end{array}
$$
\caption[]{\small The contribution of the diagrams $A_1$-$A_4$ of Fig.~\ref{fig:diaA} at $q^2 = 0$ for an on-shell photon. One observes that on a qualitative level there are four types of transitions, 
the $B$ or $D$ and charged or uncharged. 
The notation $(bD)^{0}$ for instance means a $b \to (d,s)$ transition in a charge neutral meson. 
In all cases, the charge conjugate transition follows by simply reversing the sign, since all amplitudes are proportional to the charges of the valence quarks.
Together with the non-spectator correction $G_i^{(ns)}$, this constitutes the relevant information 
for $B(D) \to V \gamma$ decays. Note $G_1^{(\perp)}(0) = G_2^{(\perp)}(0)$. Further information is given in subsection \ref{sec:HVgamma}. 
The uncertainties in the real and imaginary parts are very close and we thus refrain from quoting them separately.}\label{tab:results}
\end{table}

\begin{figure}[ht]
 \centerline{\includegraphics[width=3.2in]{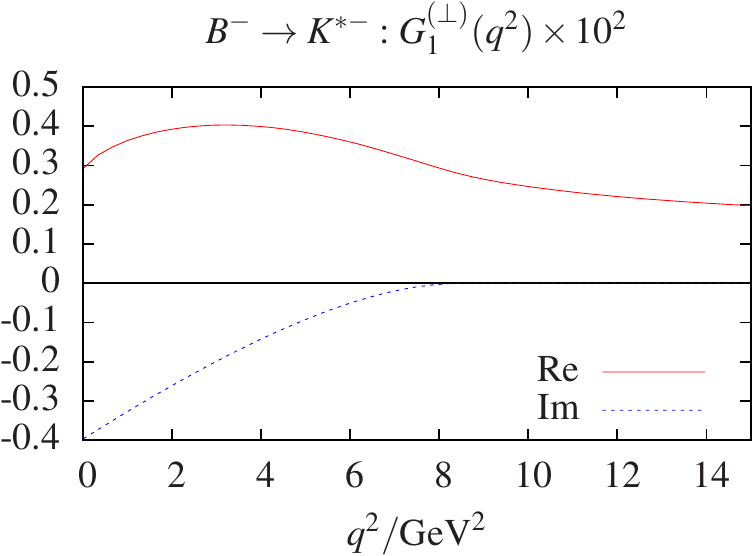}
 \includegraphics[width=3.2in]{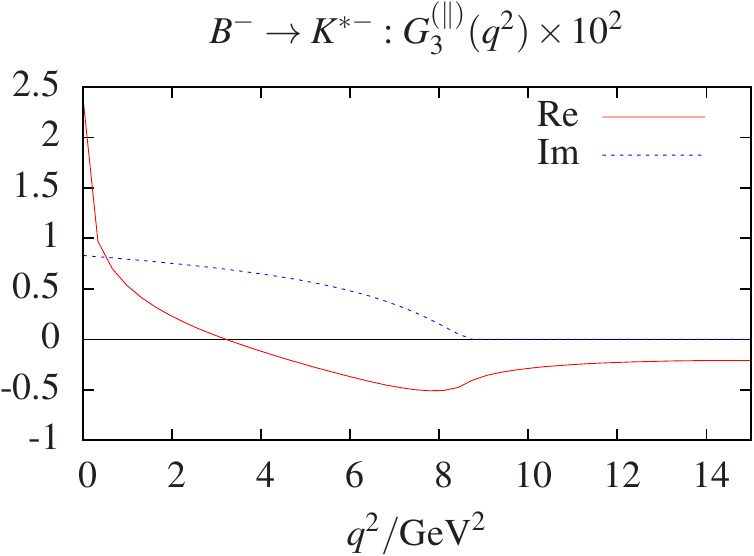}  }
\centerline{ \includegraphics[width=3.2in]{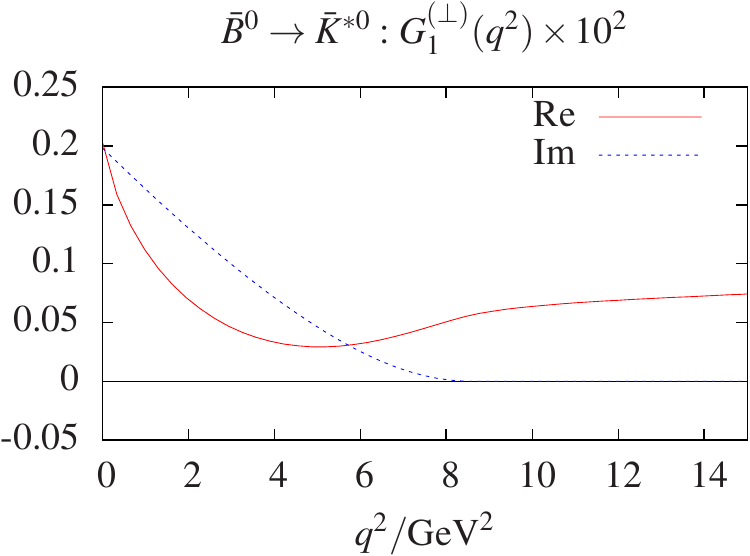}
 \includegraphics[width=3.2in]{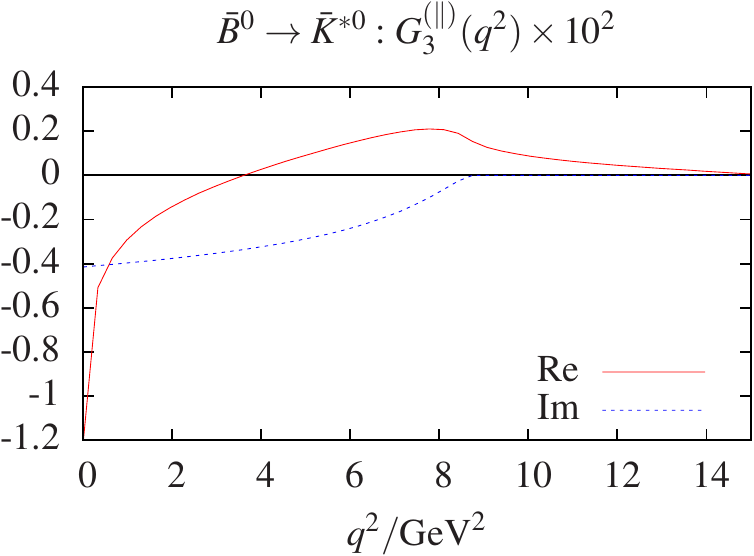}}
 \caption{\small Plots of $G^{(\perp)}_1(q^2)$ and $G_3^{(\parallel)}(q^2)$ for charged and uncharged $B$ mesons. Any other $G_\iota$-function where a $U$- or $D$-type flavour is exchanged is qualitatively  similar. 
As usual $U$- and $D$-type stand for the $u,c,t$ and $d,s,b$-flavours. For further qualitative discussion the reader is referred to subsection \ref{sec:qualitative}.}
 \label{fig:plots}
 \end{figure}

The superscripts $\{\perp,\parallel,P\}$ 
refer to the projections onto the corresponding light-meson DA e.g. \eqref{eq:DA}\footnote{Note these labels are not
necessarily in one-to-one correspondence with the amplitudes ${\cal T}_{\perp,\parallel,P}$ as used in \cite{BFS01}}.  For the sake of clarity it would be better to replace the notation by $G_i^{(\perp)} \to 
G_i|_{\phi_\perp}$  but  we shall not do so in order to retain  a compact notation. 
Out of the seven functions \eqref{eq:Gdec}, four satisfy relations so that the full function can be reconstructed 
by three of them:
 \begin{equation}
 \label{eq:allweneed}
 V: \;\; G^{(\perp)}_1(q^2), G_3^{(\parallel)}(q^2) \;, \qquad P: \;\; G_T^{(P)}(q^2) \;,
\end{equation}
The four relations required are:
 $G^{(\parallel)}_1(q^2) = 0$, $G^{(\parallel)}_2(q^2) = 0$, $G^{(\perp)}_2 = (1 - q^2/m_B^2) G^{(\perp)}_3$ and $G^{(\perp)}_2 = (1 - q^2/m_B^2) G^{(\perp)}_1$.  The third relation assures a finite decay width 
 in the limit $m_V^2 \to 0$ (as employed here) c.f. appendix \ref{app:results} and \cite{LZ12b}. 
The fourth relation is of the  large energy effective theory (LEET)-type as found for the form factors in \cite{LEET98}. 
 The latter  can be explained at this level in a straightforward way 
c.f. appendix \ref{app:results}.
Furthermore, in the ultra-relativistic approximation $m_V^2 \to 0$, the projections  
$G_T^{(P)}(q^2)$ and $ G_3^{(\parallel)}(q^2)$  are proportional to each other modulo 
a replacement of the corresponding DA c.f appendix \ref{app:results}.

For the sake of completeness, we shall give the sum rule expression for $G^{(\perp)}_1(q^2)$
\begin{eqnarray}
G^{(\perp)}_1(q^2) &=& \frac{1}{(m_B^2 f_B)|_{SR_0}} \int_{m_b^2}^{s_0}  \, e^{\frac{m_B^2-s}{M^2}} \rho^{(\perp)}_1(s) 
 \nonumber \\[0.1cm] 
\rho^{(\perp)}_1(s)  &=& p \int_0^1 du \phi_\perp(u)  \sum^d_{i = a}( b^\perp_i \rho_{B_i}(u,s) + c^\perp_i \rho_{C_i}(u,s) )\,,
\end{eqnarray}
where $p \equiv C_F (\alpha_s/4\pi) f_V^\perp m_b^2 Q_b/(-2)$, the sum runs from $a$ to $d$ alphabetically 
and $\rho_{B(C)_i}$ and $b(c)^\perp_i$ are given in Eqs.~\eqref{eq:rho} and \eqref{eq:bcperp} respectively. 

The central hadronic input parameters and their uncertainties are given in appendix \ref{app:input}. 
The collinear factorisation scale is chosen to be $\mu_F^2 = m_b(m_c) \Lambda_{\rm had} \simeq 
  m_b(m_c)\,0.8 \GeV$ for $B(D)$ transitions. This scale corresponds to the momentum 
transfer and is standard for hard-spectator contributions. 
We consider all type of FCNC $b \to (d,s)$-, $c \to u$-transitions of $B(D)$ meson into a light $V(P)$ meson 
as indicated in Tab.~\ref{tab:0}, with the exception of $P = \eta,\eta'$. This sums up to a total of $19$ transitions; 
$11$ to a vector  and $8$  to a pseudoscalar. 
Central values at $q^2 = 0$ for $G_1^{(\perp)}(0)$, as required for $B(D) \to V\gamma$-transitions (c.f.subsection \ref{sec:HVgamma}), and uncertainties  are collected in 
Tab.\ref{tab:results}. 
The validity of the $q^2$-range of our computations is discussed in subsection \ref{sec:q2-range}.  
 Let us turn to the discussion of uncertainties.
We vary the Borel parameters $M^2[G]$, $M^2[f_H]$, the continuum threshold $s_0$, the heavy quark mass $m_b$, the decay constants and the condensates as indicated in appendix \ref{app:input}. 
The major uncertainties come from varying $s_0$, $m_{h}$ and $\mu_F$ which amount 
to about $11[15]$, $8[7]$, and $5[20]$\% for $B[D]$-transitions respectively. The uncertainties in the decay constants can be significant 
depending on the final state meson as they enter linearly. We expect violation of quark-hadron duality to be accounted for by variations of $s_0$.
There are two further sources of uncertainty which are not taken care of by varying parameters.
First, the scale dependence of the operator $\tilde O_8(\mu_{\rm UV})$,\footnote{In physical processes, such as $B \to K^* \gamma$, this is compensated by the Wilson coefficients.} 
especially since we do not include proper radiative corrections in $\alpha_s$.  At 1-loop level the diagonal  anomalous dimension is $\gamma_{88} = C_F$ in conventions where $\gamma_m = 6 C_F$ and is fortunately small. Evolving at leading log level from $\mu = 1\GeV$ to $m_b$ leads to a $7\%$-effect 
which we shall adapt as an estimate of this uncertainty. 
Second, the omission of twist-3 and higher twists: 
on grounds of past experience we attribute a $15\%$ uncertainty to them. 
Note that the Borel mass is chosen to suppress the latter, yet keeping violations of quark-hadron duality acceptably small, as explained in appendix \ref{app:input}. 
Finally all the parametric variations, as described above, and the uncertainty of higher twist and $\mu_{\rm UV}$ are added in quadrature, as we do not see a reason for strong correlations. The final uncertainties
along the central values are collected in  Tab.~\ref{tab:results}. 

\subsection{Qualitative discussion}
\label{sec:qualitative}

As discussed in the caption of Tab.~\ref{tab:results} there are four qualitatively different transitions
depending on whether the initial meson is either of $b$ or $c$ flavour and on whether it is charged or not, which is of course
a manifestation of the sensitivity to isospin. The $b$-types are plotted in Fig.~\ref{fig:plots}. 
The $q^2$-dependence  is somewhat more complex than the one of an ordinary form factor 
$B \to \pi$. In the latter case the $q^2$-dependence is merely governed by a series of poles, starting at  $q^2 = m_{B^*}^2$, 
and higher multihadron cuts.  For this reason fitting that form factor is rather simple. In our case at hand, as discussed in the next subsection, the photon couples 
to all kinds of flavours and thus poles in $q^2 = m_\rho^2,  m_{B^*}^2, \Upsilon(\bar bb)$ appear.
Furthermore there are genuine LD contributions which result in strong phases for $q^2, P^2 > 0$ as 
discussed in subsection \ref{sec:strong} and illustrated in  Fig.\ref{fig:LD}. Moreover we note that the imaginary part decreases with $q^2$. This is to be expected as the process shown in Fig.~\ref{fig:LD}(left) is more and more
off-shell for higher $q^2$, at least at leading order $\alpha_s$.

In Tab.~\ref{tab:ratios} we reproduce values for $G_1(0)$
for the spectator contributions  $G_1^{(s)}(0)$, the non-spectator contributions  
$G_1^{(ns)}(0)$, their sum $G_1(0) = G_1^{(s)}(0)+ G_1^{(ns)}(0)$
as well as ratios between the latter and the SD penguin form factors $T_1(0)$.
Let us briefly discuss the heavy quark scaling of the various parts\footnote{A word of caution seems appropriate here. 
In section \ref{sec:hqlimit} it is found  that, for  diagrams $A_1$ and $A_2$, the  leading heavy quark term, including a non-expandable logarithm in $m_b$, gives roughly $50\%$ of the contribution.  
Whereas this points towards  large corrections, it does not imply that qualitative behaviour cannot  be understood from the leading scaling.}. From $T_1(0) \sim m_b^{-3/2}$ (as first derived in \cite{CZ90})
it follows that  $G_1^{(ns)}(0) \sim m_b^{-3/2}$ from the formulae given in appendix \ref{app:non-spectator}. For $G_1^{(s)}$ it is useful to split the matrix elements
according to whether or not the photon is emitted from the spectator: 
\begin{equation}
\label{eq:Qdec}
G_1^{(s)}(0) = Q_h G_1^{h, (s)}(0) +  Q_q G_1^{q, (s)}(0)  \;, \qquad h \in \{b,c\} \;, q \in \{u,d,s\}
\end{equation}
The discussion of section \ref{sec:hqlimit} suggests that:  $G_1^{h,(s)}(0) \sim m_b^{-3/2}$ 
and $G_1^{q,(s)}(0) \sim m_b^{-5/2}( \ln m_b + {\cal O}(1))$.
Let us discuss the numerical ratios.
The ratios of $|G_1^{(\perp)}(0)/G_1^{(ns)}|$ are between $20\%$ and $59\%$ and vary considerably according 
to the charge and flavour of the heavy initial meson.  
The ratio of  $|G_1^{(s)}(0)/ T_1(0)|$ is around 
$2\%$ for the $B$ meson and considerably larger for the $D^{0(-)}$ at $5\%(13\%)$.
The ratio of the total $G_1(0)$ to the SD part, $|G_1(0)/ T_1(0)|$, is $7\%$ for the 
$B$ meson and rather sizeable for the $D^{0(-)}$: $21\%(34\%)$.
An interesting aspect is the comparison of the $B$ and $D$ matrix elements themselves. 
To obtain a  meaningful answer  we have to use the decomposition \eqref{eq:Qdec}:
\begin{equation}
\label{eq:R}
R_h = \frac{ G_1^{b, (\perp)}(0) [B \to \rho  \gamma] }{G_1^{c, (\perp)}(0) [D \to \rho  \gamma]  } =0.14 \;, \quad 
R_l = \frac{ G_1^{q, (\perp)}(0) [B \to \rho  \gamma] }{G_1^{q, (\perp)}(0) [D \to \rho  \gamma]  } =0.05 +0.04 i \;.
\end{equation}
Using the scaling behaviour  above we would infer that 
$$|R_{h[q]}| \simeq  \alpha_s(\sqrt{m_c \Lambda_{\rm had}})/   \alpha_s(\sqrt{m_b \Lambda_{\rm had}}) (m_c/m_b)^{3/2[5/2]} \simeq 0.2[0.06]$$ which 
which is very close to the values in Eq.~\eqref{eq:R}.

\begin{table}[H]
\addtolength{\arraycolsep}{3pt}
\renewcommand{\arraystretch}{1.2}
$$
\begin{array}{ l  |  rrrr } 
\text{type}   & B^- \to \rho^- \gamma & \bar B^0 \to \rho^0 \gamma & D^+ \to \rho^+ \gamma & D^0 \to \rho^0 \gamma \\ \hline
G_1^{(s)}(0)  \cdot 10^{-2}  &   0.29-0.39i &  0.22+0.19 i &  -1.9+2.5  i &  -7.0-5.0i  \\
G_1^{(ns)}(0)  \cdot 10^{-2}   &  0.90+ 1.3 i   &   0.90+ 1.3 i  & -8.5 - 12 i & -8.5 - 12 i  \\ 
G_1(0)   \cdot 10^{-2} &   1.2+0.91 i & 1.1 + 1.5 i & -10 -9.5 i & -16-17 i \\ \hline
 \left| G_1^{(s)}(0)/G_1^{(ns)}(0)  \right| [\%] & 31 & 18 & 21  & 58 \\
 \left| G_1^{(s)}(0)/ T_1(0) \right| [\%] & 2  & 1 &  4 & 12     \\
   \left| G_1(0)/ T_1(0) \right| [\%] &  6  & 7  &  20 & 33
\end{array}
$$  
\addtolength{\arraycolsep}{-3pt}
\renewcommand{\arraystretch}{1}
\caption[]{\small  Comparison of various parts of the four characteristic types of  $G_\iota$-functions. 
See subsection \ref{sec:qualitative} for comments.  
 For the $T_1(0)$ form factors we use 
$T_1^{B \to \rho}(0) = 0.27$ \cite{BZBtoV} for $B\rightarrow\rho$ and $T_1^{D \to \rho}(0)= 0.7$, e.g. \cite{Wu:2006rd}, for $D\rightarrow\rho$ as reference values. 
Note $ G_1^{(s)}(0) =  G_1^{(\perp)}(0)$ at our level of twist-approximation. 
The ratio of $G_1^{(ns)}$ to $T_1(0)$ can directly be inferred from the formula 
\eqref{eq:G1nstoT1}.} \label{tab:ratios}
\end{table}

\subsection{Validity of computation in $q^2$-range}
\label{sec:q2-range}

 Let us discuss the validity of our computation in the $q^2$-range in some more detail than in section 
 \ref{sec:strong}.  The computation cannot 
 be trusted when either real QCD or perturbative QCD, as employed here\footnote{By which we mean the LC-OPE with perturbatively computed hard scattering kernels.}, predicts the production of particles,
which would be hadrons and quarks \& gluons in the respective cases.
This happens in real QCD when $q^2$ reaches the  $\rho$-, $B^*_{d,s}$- and $\Upsilon(\bar bb)$ thresholds for $J^{PC} = 1^{--}$-mesons. The corresponding  production thresholds for perturbative QCD are of  the 
two-valence quark-type and occur at $q^2$:  $(2 m_q)^2$, $(m_b+m_{d,s})^2$
 and $(2m_b^2)$ respectively 
 
As previously stated  the $\rho$-threshold leads to the exclusion of the region 
 $0 < q^2 < (\simeq 1\,\GeV^2)$ for $B \to V ll$. The quark threshold at  $(m_b+m_{d,s})^2$ indicates that
 the LC-OPE is not valid a few $\rm GeV$ below that value. This is the case for all diagrams except 
 $A_1-A_2$ which do not have these thresholds and therefore the validity ought to extend a few $\GeV$ 
 below $B^*$-resonance and thus basically to the endpoint of the physical region.

\subsection{Summary for $B(D) \to V \gamma$ }
\label{sec:HVgamma}

For the reader's convenience, we briefly summarise the essentials points
for $B(D) \to V \gamma$ decay.
\begin{eqnarray}
\label{eq:HVgamma}
B(D) \to V \gamma: \quad  G_1(0) &=& G_2(0) = G_1^{(\perp)}(0) + G_1^{(ns)}(0)  
\end{eqnarray}
with 
\begin{equation}
 G_1^{(ns)}(0)\stackrel{\eqref{eq:Gns}}{=}     \left( \frac{3 \alpha_s(m_h)}{4 \pi} \right) Q_h F_8^{(7)} T_1(0) \;, 
\label{eq:G1nstoT1}
\end{equation} 
where $h = b(c)$, $Q_{b(c)} = -1/3(2/3)$ and $F_8^{(7)}$ are taken from Ref~\cite{AAGW01}.
The generic amplitude assumes the following form:\footnote{The amplitudes ${\cal A}_{1,2}$ up to normalisation are often denoted by $A_{\rm PC, PV}$ in the literature.} 
\begin{equation}
\label{eq:AHVgamma}
{\cal A}(B(D) \to V \gamma) \sim  \Big(  {\cal A}_1 (P_1\! \cdot \! \epsilon)  +   {\cal A}_2  (P_2\! \cdot \! \epsilon)  \Big) \;,
\end{equation}
where ${\cal A}_{L,R} =  {\cal A}_1 \pm {\cal A}_2$ correspond to left- and right-handed photon polarisations.
Our result and the  leading SD penguin read
\begin{equation}
{\cal A}_1 =  {\cal A}_2 = C_7 T_1(0)  + C_8 G_1(0) + ..  \;.
\end{equation}
Using the notation  ${\cal O}_{7,8}'  =   {\cal O}_{7,8}|_{\gamma_5 \to -\gamma_5}$  for the penguin operators with opposite chirality and the corresponding Wilson coefficients one gets:
\begin{equation}
{\cal A}_{1,2}  = C_7 T_1(0)  + C_8 G_1(0) \pm  ( C'_7 T_1(0)  +  C_8' G_1(0)) +  ..  \;,
\end{equation}
where we have used $T_1(0)= T_2(0)$ and $G_1(0) = G_2(0)$. The former is an equality and the latter 
is a result of our leading twist-2 computation.

\section{Comparison with QCD factorisation}
\label{sec:QCDF}

\begin{figure}[h]
 \centerline{\includegraphics[width=6.0in]{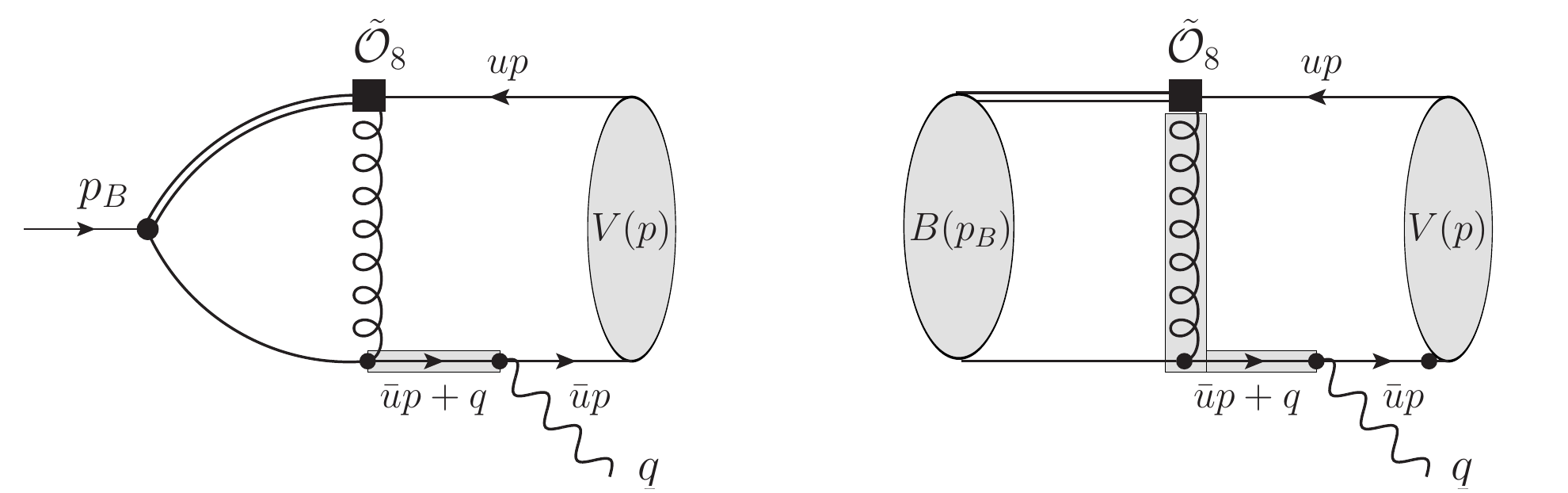}}
 \caption{\small The shaded propagators that scale like $1/(\bar u m_B^2)$ in both figures. 
 (left) Diagram of LCSR or the LC-OPE respectively (right) Diagram in QCDF. 
 Thus $x_\perp^{QCDF} \sim 1/\bar u^2$ and $x_\perp^{LCSR} \sim \ln (\bar u)/\bar u$ at worst, as explained in the text.}
 \label{fig:endpoint}
 \end{figure}

In this section we shall compare our results with QCDF \cite{KN01}.
More precisely the diagrams $A_1$ and $A_2$\footnote{Note the sum of these two diagrams is well-defined 
as they constitute the contribution proportional to the spectator charge.}, in Fig.~\ref{fig:diaA},  at $q^2 =0$ 
corresponding to $Q_q G_1^{q,(s)}(0)$ \eqref{eq:Gdec} shall be considered where the formulae take on a rather simple form. 
Let us first define the quantities in question and then point towards the points we would like
to investigate.
We parameterise the $G_1$-function at $q^2=0$ as follows:
\begin{equation}
\label{eq:KNO8}
G_1(0) = \Big[  \underbrace{ \frac{\alpha_s}{4 \pi} \frac{C_F}{N_c}  12 \pi^2   \frac{f_\perp f_B}{m_B^2}   }_{\sim \, m_b^{-5/2}}  \Big] ( Q_q X_\perp  + Q_b  \overline X_\perp)   \;,  \quad\end{equation}
with $X_\perp$ as in Ref.~\cite{KN01},
\begin{equation}
\label{eq:Xperp}
 X_\perp \quad = \quad  \int_0^1  \phi_\perp(u) x_\perp(u) \;,
\end{equation}
\begin{equation}
\label{eq:XQCDF}
 x_\perp^{QCDF}(u) \quad =\quad   \frac{1+\bar u}{3 \bar u^2}  
\end{equation} 
and likewise for the quantity $\overline X_\perp$. The LCSR result in this limit reads:
\begin{equation}
\label{eq:XLCSR}
 x_\perp^{LCSR}(u)  \quad = 
\int_{m_b^2}^{s_0} ds \, e^{\frac{m_B^2-s}{M^2}} \rho(s,u)   \;,
\end{equation}
with
\begin{eqnarray}
\label{eq:rho0}
\rho(s,u) &=& \underbrace{\frac{m_b^2 N_c}{12\pi^2 f_B^2}}_{\equiv c  m_b^3} \Bigg[
\frac{\log\left(\frac{\bar us(m_b^2 + P^2 - s)}{P^2(m_b^2 - us)}\right)}{P^2 - \bar us} - \frac{s - m_b^2}{\bar u s P^2} \Bigg] \;, \nonumber \\[0.1cm]
\bar \rho(s,u) &= & \frac{m_b^2 N_c}{12\pi^2 f_B^2}\Bigg[ -  \left(\frac{s-m_b^2}{u s P^2} -  \theta(us-m_b^2)\left(\frac{us-m_b^2}{2u^2sP^2} + \frac{\log\left(\frac{us}{m_b^2}\right)}{2uP^2}\right)\right)
\Bigg] \;.
\end{eqnarray}
We would like to emphasise that we have computed the result in 
Eq.~\eqref{eq:XQCDF} anew and that we have found agreement with reference \cite{KN01}.
We have kept the contributions of diagrams $A_{3,4}$, which correspond to $\overline X_\perp$, in the expression 
above since their large $m_b$-behaviour is interesting per se.
A few remarks about the $m_b$-behaviour  are in order.
The term in the bracket in Eq.~\eqref{eq:KNO8} scales as $m_b^{-5/2}$, taking 
into account $f_B \sim m_b^{-1/2}$.
The coefficient $c$ in Eq.\eqref{eq:rho0} is  $ {\cal O}(m_b^0)$. 
 The expression $X^{QCDF}_\perp$ is ${\cal O}(1)$.
The questions we would like to investigate are:
\begin{itemize}
\item[a)] The presence and absence of an endpoint divergence at leading order $\alpha_s$, for $\bar u \to 0$,  
in $X_\perp^{QCDF}$ and $X_\perp^{LCSR}$ respectively. 
\item[b)] In what respect $X_\perp^{QCDF}$ and $X_\perp^{LCSR}$ can be compared 
to each other.
\item[c)] The absence and presence of an imaginary part, at leading order in $\alpha_s$,  in $X_\perp^{QCDF}$ and $X_\perp^{LCSR}$ respectively.
\end{itemize}
The answers to these questions are, certainly, tied to each other. 
We shall begin by discussing question a).  Assuming the usual 
endpoint behaviour\footnote{This is true to any finite order
in the Gegenbauer expansion. Since the Gegenbauer polynomials are a complete set 
on the $[0,1]$-interval this could be changed by an infinite sum of them. 
This is not the currently accepted scenario. 
},
\begin{equation}
\label{eq:endpoint}
 \phi_\perp(u) \stackrel{ u \simeq 1}{\to} 6 \bar u u \;,
 \end{equation}
  the most singular part in \eqref{eq:XQCDF}, 
  \begin{equation}
 \label{eq:endpoint_div}
 x_\perp^{QCDF} = \frac{1}{3 \bar u^2} + {\cal O}(\bar u^{-1})  \quad \Rightarrow  \quad 
 X_\perp^{QCDF} =  2 \int_0^1 \frac{du}{ \bar u} + \text{finite}
 \quad 
 \end{equation}
 convoluted as in \eqref{eq:Xperp} with \eqref{eq:endpoint}
 leads to  logarithmic endpoint divergence\footnote{We note that these divergences 
are also regulated by $q^2 \neq 0$ as they originate 
from $(\bar u p + q)^2 = u q^2 - \bar u u p^2  + \bar u (p+q)^2  \to \bar u ( m_B^2 - u m_{V}^2) +  u q^2 $ but not by a finite meson final state mass.}.
 The endpoint configuration $u \simeq 1$ corresponds to the situation where the non-spectator quark carries all
the momentum.
On a purely technical level the divergent integral  arises from the fact that two propagators 
assume the same form $1/(\bar u m_B^2)$, c.f. Fig.\ref{fig:endpoint}(left), 
 as the momentum fraction of the spectator quark is neglected due to $\Lambda_{\rm QCD}/m_b$ suppression.  
In view of this and potential transverse corrections it was advertised  in \cite{BBNS}, that for $B \to \pi\pi$ and similar cases the replacement 
$1/(\bar u m_B^2) \to 1/((\bar u+ \epsilon) m_B^2)$ should be made ($\epsilon = \Lambda_h/m_b$ with $\Lambda_h$ some hadronic scale of the order of the QCD-scale) 
and  a correction term included to account
for missing soft contributions with possible strong phases.
The endpoint divergent integral in \eqref{eq:endpoint_div} becomes,
\begin{eqnarray}
\label{eq:remedi1}
x_\perp^{QCDF} &\to& (1+ \rho e^{i \phi})  \Theta\left( \bar u - \frac{\Lambda_h}{m_b} \right) \,  \frac{1}{3 \bar u^2}  + {\cal O}(\bar u^{-1})  
 \\[0.1cm]
 \label{eq:remedi2}
\Rightarrow  \quad 
 X_\perp^{QCDF} &=&  
 2  (1+ \rho e^{i \phi})  \,  \ln \left(  \frac{m_b}{\Lambda_h} \right) + \Lambda_h\text{-independent} \;,
\end{eqnarray}
with $\rho \in [0,1]$ and $\phi \in [0,2\pi]$ being numbers parametrising the above mentioned  corrections. 
Thus  changes can be expected if the heavy quark limit is not assumed  as is the case in LCSR. Yet 
the question we would like to address 
is whether there are qualitative differences beyond the behaviour of the RHS in  Eqs.~(\ref{eq:remedi1},
\ref{eq:remedi2}).

In the LCSR computation there is only one  propagator with manifest $1/(\bar u m_B^2)$-behaviour, c.f. Fig.~\ref{fig:endpoint}. 
Thus the question: Is there another one hidden in the loop? The answer to that
is no as it would correspond to a power IR-divergence whereas it is known that in 
four dimensions IR-singularities, be they soft or collinear, are at worst logarithmic in 
nature, e.g. \cite{Muta}. The smoother behaviour 
of the diagram in Fig.~\ref{fig:endpoint}(left)  with respect to the QCDF result Fig.~\ref{fig:endpoint}(right) is in line with 
the improved IR-behaviour of inclusive processes as manifested  in the classic IR-cancellation theorems 
of the  Bloch-Nordsieck and Kinoshita-Lee-Nauenberg  type \cite{Muta}\footnote{At this point it is more inclusive because we sum over all states with $B$-meson quantum numbers and because 
there are additional LD-contributions Fig.~\ref{fig:LD}(left). The former will be removed once 
the correlation function is inserted into modified dispersion integral \eqref{eq:SR2}.
It remains to be investigated what happens when the $m_b$-scaling 
of $s_0$, $m_B$ and $M$ is made explicit as  done in subsection \ref{sec:hqlimit}.}.
Inspection of the graph  Fig.~\ref{fig:endpoint} reveals that there can at most
be a collinear divergence in the limit $\bar u \to 0$ and $p^2 = q^2 = 0$.
The potential endpoint sensitive terms are parametrised as follows, 
\begin{equation}
\label{eq:endpoint_finite}
x_\perp^{LCSR} \sim \alpha_\perp \,  \frac{\ln (\bar u)}{\bar u} + \beta_\perp \ln (\bar u) + \gamma_\perp \, \frac{1}{\bar u} \;.
\end{equation}
Note that they are all integrable assuming the DA Eq.\eqref{eq:endpoint}.
From  Eq.~\eqref{eq:rho0}\footnote{Integration over $ds$ is not going to change anyting at this point.} it is found that: $\alpha_\perp =0$, $\beta_\perp \neq 0$, $\gamma_\perp \neq 0 $.
The absence of the most singular term  $  \ln ( \bar u) /\bar u$ appears to be accidental; such terms are present in the $P/V^\parallel$-contribution.
In summary, the endpoint behaviour  of the $x_\perp^{LCSR}$ \eqref{eq:endpoint_finite} differs 
from $x_\perp^{QCDF}$ \eqref{eq:endpoint_div} even when finite $m_b$-effects are added by hand 
\eqref{eq:remedi1}.

Before attempting an interpretation of this difference we should try to reflect on question b), namely to what degree it makes sense to compare the QCDF and the LCSR result at face value.

 We advocate that, within the approximations, the QCDF contribution is contained in the LCSR result 
 but the converse is not true.
 For example  the gluon in Fig. \ref{fig:endpoint}(right) is not necessarily  the hard gluon of QCDF but can also
be a gluon that hadronises into a $3$-particle $(qs)_{0^\pm}$-state  as shown 
in Fig~\ref{fig:LD}(left). Moreover there are cuts of the $3$-particle type for the $B$-meson as well, c.f. 
Fig.~\ref{fig:parasite}.
Possibly it is helpful, at this point, 
to note that  there is a crucial difference between the two approaches. In QCDF one computes 
a specific sub-process and the corresponding scaling of the momenta 
leads to a clear physical picture of the dynamics of that sub-process,
whereas in LCSR one computes a correlation function, in a domain 
where it is believed to be valid, and extracts the matrix element by suitable methods
such as dispersion relation and Borel transformation.  
Thus the physical parton configurations are, generically, not immediately deducible from the correlation function.  

In summary the LCSR result is not endpoint divergent,
yet sensitive to the endpoint\footnote{At leading order in $\alpha_s$ the most sensitive terms is 
$\Delta = \int 6 u \bar u \frac{1}{\bar u} = 3 \sum_{n \geq 0} (-1)^n a_n$ 
where $a_n$ are the Gegenbauer moments e.g. \cite{CK00}.  Explicit computations of the Gegenbauer moments as well as an investigation of the pion form factor \cite{BKM99} show that 
the influence of the Gegenbauer moments on this quantity is rather moderate (at the 10-20\% level).}.
We have seen that the amendment \eqref{eq:remedi1} is not enough to obtain 
a similar qualitative behaviour of  $x^{QCDF}_\perp$ and $x^{LCSR}_\perp$. 
Whether or not this is due to the fact that $x^{LCSR}_\perp$ constitutes in addition to the
physics present in $x^{QCDF}_\perp$ a  LD-part Fig.~\ref{fig:LD}(left) is a question that we did not address. 
The question of why the QCDF contribution does not admit, in its current form, a heavy quark expansion 
can be illuminated by investigating what happens when a LCSR heavy quark expansion is attempted. 
This is the goal of the next subsection.

\subsection{Heavy quark limit and the dependence on the value of $m_b$}
\label{sec:hqlimit}

\begin{figure}[h]
 \centerline{
\includegraphics[width=3.0in]{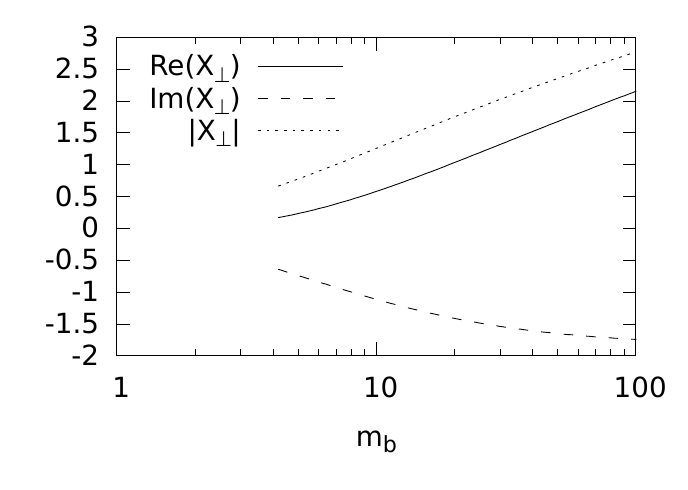} \quad 
  \includegraphics[width=3.0in]{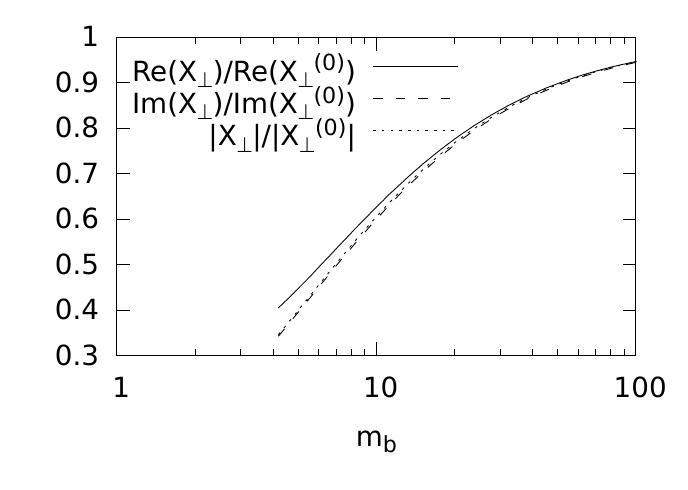}
 }
 \caption{\small (left)  Absolute value, real and imaginary part of $X_\perp^{LSCR}$ 
 as a function of $m_b$ assuming the rescaling \eqref{eq:scaling}. The plot makes it apparent 
 that the hierarchy of the real and imaginary part is rather dependent on the actual value 
 of $m_b$.  (right) Ratio of the asymptotic expression $(X_\perp^{LCSR})^{(0)}$ in \eqref{eq:XLCSR0} over the expression including all $m_b$ corrections within the 
 rescaling \eqref{eq:scaling}. Note non-leading order corrections decrease the quantity $X_\perp$.}
 \label{fig:hq-limit}
 \end{figure}

In this section we would like to investigate whether the two approaches 
behave similarly in the heavy quark limit.
Although this cannot be done in a absolutely transparent way, 
a rescaling  in the heavy quark mass\footnote{We refrain from rescaling
$f_B \to (f_B)_{\rm stat} m_b^{-1/2}$. We shall simply use this known scaling behaviour in what 
follows.} $m_b$ has been proposed in \cite{CZ90,BBBD92}:
\begin{equation}
\label{eq:scaling}
m_B \to  m_b + \bar\Lambda \;, \quad 
s_0 \to  m_b^2 + 2m_b\omega_0 \;, \quad 
M^2 \to  2m_b\tau \;,  
\end{equation}
where $\bar \Lambda$, $\omega_0$ and $\tau$ are all hadronic scales of which 
$\bar \Lambda$ is, of course, rather well-known. 
 In many cases this has reproduced 
the leading order behaviour from a proper heavy quark treatment of the same quantity.
The expansion in $m_B$ and $s_0$ are of leading order 
and the Borel mass $M^2$ is adjusted such that the exponential is free of 
powers of $m_b$.
The expression $x^{LCSR}$ can then be rewritten in terms of the dimensionless
integration variable $z$:
\begin{equation}
\label{eq:xLCSR}
x^{LCSR}_\perp(u)  = 2 m_b \omega_0 \int_0^1  e^{ \frac{ (\bar \Lambda -   \omega_0 z )}{\tau}} \rho(m_b^2 + 2 m_b \omega_0 z,u) dz \;.
\end{equation} 
Using the asymptotic DA $\phi_\perp(u) = 6 u \bar u $ in \eqref{eq:Xperp}, integrating over $du$
and isolating a non-expandable logarithm  
we get:
\begin{eqnarray}
X_\perp^{LCSR} &=&  \underbrace{\left[\frac{N_c\omega_0^2 }{f_B^2\pi^2}\right] \Bigg\{
\frac{2\omega_0}{m_b}\left(\left(\ln \left(\frac{m_b}{2 \omega_0}\right)\!-\! i \pi\right)\aver{z^2}  \!-\! \aver{z^2\ln z}\right)}_{(X_\perp^{LCSR})^{(0)}} + {\cal O}\left(\frac{\Lambda_{\rm QCD}^2}{m_b^2} \right) \Bigg\} \nonumber \\[0.1cm]
\overline X_\perp^{LCSR}  &=&\left[\frac{N_c\omega_0^2 }{f_B^2\pi^2}\right]  \Bigg\{ \left( \langle z\rangle\left(\frac{2\bar\Lambda}{m_b}\!-\!  1 \right)\!+\! \frac{2\omega_0}{m_b}\langle z^2\rangle\right)  + {\cal O}\left(\frac{\Lambda_{\rm QCD}^2}{m_b^2} \right) \Bigg\}
\label{eq:XLCSR0}
\end{eqnarray}
with $ \aver{f(z)} = \int_0^1 e^{ \frac{ (\bar \Lambda -   \omega_0 z )}{\tau}}  f(z)   dz  $ 
being a number of order one.
A few remarks are in order:
\begin{itemize}
\item
From the appearance of the imaginary part at leading order it would seem
in the heavy quark limit \eqref{eq:scaling}  that
the QCDF and LCSR computations cannot
 be compared as the former are real.  
This would suggest that the LD contributions c.f. Fig.~\ref{fig:LD}(left), 
responsible for the CP-even phases, do not seem 
to be suppressed in the heavy quark limit for spectator emission. 
\item Eq.~\eqref{eq:XLCSR0} suggests, using the notation as in Eq.~\eqref{eq:Qdec}, that 
$$G_1^b(0) \sim m_b^{-3/2}\;, \quad  
G_1^q(0) \sim m_b^{-5/2}( \ln m_b + {\cal O}(1))$$
These  scaling behaviours are in line with  Refs.\cite{BB01,BFS01} for $G_1^b(0)$ and  Ref.\cite{KN01} for $G_1^q(0)$. The endpoint divergence can be associated with the non-expandable logarithm to be discussed below.
\item 
The $\ln m_b$ term signals that the result, using the rescaling \eqref{eq:scaling}, 
is not expandable in powers of $1/m_b$.  This statement is of course dependent 
on the behaviour of the DA at the endpoint $u \simeq 1$ \eqref{eq:endpoint}.
This can be further illustrated by first expanding the density $\rho$ in Eq.~\eqref{eq:xLCSR} 
in inverse powers of the heavy quark mass.
To leading order we get,
\begin{alignat}{2}
\text{Re}[\rho] &= \frac{2 c \omega_0^2 z^2}{m_b}\frac{1+\bar u}{\bar u^2}  \;, \quad 
&\text{Im}[\rho] =& -\frac{cm_b\pi}{u}\theta\left(u-\left(1-\frac{2\omega_0 z}{m_b}\right)\right) \nonumber \\
\text{Re}[\bar \rho] &= - \frac{2c\omega_0 z}{u}\left(1-\frac{2\bar\Lambda}{m_b}-\frac{2\omega_0}{m_b}z\right) \;, \quad &\text{Im}[\bar \rho] =& 0 \;,
\end{alignat}
up to order ${\cal O}(\Lambda_{\rm QCD}^2/m_b^2)$.
Thus one recovers the endpoint singularity of the QCDF-result. 
Note, the difference in powers of $m_b$ and $z$ in the real and imaginary parts is only apparent
or compensated by  the narrowness of the resulting $du$ integration interval. 
Further expansion in powers of $m_b$ in the real part leads to more and more
 endpoint divergent expression:  ${\rm Re}[\rho] \sim \frac{1}{m_b^n} \frac{1}{\bar u^{n+1}}$. 
This originates from the term $u s - m_b^2$ in the logarithm in Eq.~\eqref{eq:rho0}.

\item The rescaling \eqref{eq:scaling} allows us to investigate the numerical dependence 
of the real and imaginary parts on the mass $m_b$. As can be inferred from Fig.~\ref{fig:hq-limit}(left) 
the smallness of the real part with respect to the imaginary part at $m_b \simeq 4.6 \GeV$ is rather accidental. 

\item Information on the convergence of the $1/m_b$-expansion can be inferred from 
Fig.~\ref{fig:hq-limit}(right), though the cautionary remarks above and below equation \eqref{eq:scaling}
should be kept in mind.

\end{itemize}

\section{Summary \& conclusions}
\label{sec:conclusions}

In this work we have reported on the computation of ${\cal O}_8$ matrix elements 
between heavy pseudoscalar $B$ and $D$ states and a light vector and pseudoscalar state
and an off-shell photon by using the method of LCSR at leading twist-2 and leading $\alpha_s$. 
We have defined scalar functions of the photon momentum invariant 
$G_{1,2,3}(q^2)$ and $G_T(q^2)$ Eqs.~(\ref{eq:matel},\ref{eq:FFdec}) such that they parallel the
well-known penguin tensor form factors $T_{1,2,3}(q^2)$ and $f_T(q^2)$ c.f. Eq.~\eqref{eq:analogue}.
Central values for all flavour transitions, with the exception of $\eta$ and $\eta'$, are presented 
in Tab.~\ref{tab:results} as well as plots of the four characteristic cases in Fig.~\ref{fig:plots} 
are presented in section \ref{sec:numerics}.  
A remarkable feature is the large CP-even (strong) phase for which we give a LD interpretation  in section \ref{sec:strong} (c.f. Fig.~\ref{fig:LD}). 
This fact as well as the plots make it clear why we refer to $G_\iota(q^2)$ as matrix elements 
rather than form factors.
Comparison of    various contributions
such as spectator, non-spectator, and SD penguin photon emission can be found in Tab.~\ref{tab:ratios}.
Let us note that the $G_\iota(q^2)$-functions  are relevant for asymmetries 
of isospin- \cite{LZ12b} and CP-type (depending on new weak phases) \cite{LZ12a} rather than branching ratios.

In section \ref{sec:QCDF} we compare our computation with QCDF. The comparison is not straightforward 
as the LCSR contrary to QCDF are not tailored around a heavy quark expansion and second 
LCSR contain LD contributions of the type shown in Fig.~\ref{fig:LD}(left) which are not present in leading order QCDF.  The LCSR computation does not suffer from endpoint divergences which we trace back to the fact that IR-divergences are at worst logarithmic in four dimensions.  
When a heavy quark extrapolation of the LCSR result is attempted, c.f. section \ref{sec:hqlimit}, 
a logarithm of $m_b$ appears which might be taken as an indication  towards potential difficulties 
of the $m_b$-expansion e.g. endpoint divergences\footnote{When in the same limit  the density of the collinear momentum fraction is expanded in $1/m_b$ then indeed the same behaviour as in QCDF is found. 
It is worthwhile that qualitative differences between the two approaches remain even in that case for reason mentioned above.}.  
Whether or not an approach can be devised to deal with these endpoint divergences in the heavy quark limit is an interesting problem per se. Recent approaches known under the names of  collinear anomaly \cite{ca} 
and rapidity renormalization group \cite{rrg} might give rise to further developments leading to a consistent 
treatment of endpoint divergences in the heavy quark limit.

A remarkable feature on the technical side of our computation is the appearance of a complex anomalous threshold 
on the physical Riemann sheet for which we give various viewpoints and derivations in appendix \ref{app:analytic}.  The anomalous threshold is associated, in the three point-function, with all three propagators being on the mass shell and therefore is not related to the intermediate $B$-meson state. 
The crucial point, for the physics, is the anomalous thresholds is
well isolated from the $m_B$-pole.  This results in an exponential as well as oscillatory 
suppression by the Borel parameter such that  the extraction of the matrix element is not affected considerably.

 We shall add a paragraph contemplating on the size of the isospin asymmetry 
in $ b \to q \gamma$ due to ${\cal O}_8$, interfering with the leading ${\cal O}_7$-part, in the inclusive and exclusive case. 
In the former this was investigated in \cite{LNP06} by means of a vaccum saturation approximation and it is found that,
\begin{equation}
\label{eq:inO8} 
a_I^{\bar 0 -}(X_s \gamma)|_{{\cal O}_8} = \frac{\Gamma(\bar B^0  \to X_s \gamma) -  \Gamma(B^- \to X_s \gamma)}{\Gamma(\bar B^0 \to X_s \gamma) + \Gamma(B^- \to X_s \gamma) } = -0.05 \left( \frac{0.5 \GeV}{\lambda_B} \right)^2 \;.
\end{equation}
The symbol  $\lambda_B$ corresponds  to the first inverse moment of the $B$-meson DA whose uncertainty leads to the authors of \cite{LNP06} to attribute a spread of  $-0.02$ to $-0.19$  to \eqref{eq:inO8}. For the exlusive case we find, using our work, 
\begin{equation}
\label{eq:exO8} 
a_I^{\bar 0 -}(K^* \gamma)|_{{\cal O}_8}  = \frac{C_8 {\rm Re} [ Q_d G_1^{\bar B^0\rightarrow\bar K^{*0}\gamma }(0)- Q_u G_1^{B^-\rightarrow K^{*-}\gamma }(0)]}{C_7 T_1^{B \to K^*\gamma}(0)} =  - 0.004(2) \;.
\end{equation}
We have used $G_1(0)$ from Tab.\ref{tab:results}, $T_1(0) \simeq 0.33$ \cite{BZBtoV}, $C_7 \simeq -0.36$ and $C_8 \simeq -0.16$ \cite{AAGW01}. 
It is noted that the sign of the effect is the same but the estimate of the inclusive case is somewhat higher even given the uncertainty.  Since experimentally the inclusive rate is a 
sum of exclusive rates, the numbers in  \eqref{eq:inO8} \eqref{eq:exO8} indicate that higher states than the $K^*$ in the spectrum are more prone to isospin violation originating  from ${\cal O}_8$. At last it might be of interest to quote the current experimental averages \cite{HFAG12} $a_I^{\bar 0 -}(X_s \gamma) = 
-0.01(6)$ and $a_I^{\bar 0 -}(K^* \gamma) = 0.052(26)$. The isospin asymmetry 
in $B \to K^* \gamma$ is dominated by weak annihilation (c.a. 5\%) in the SM \cite{KN01} 
and from \eqref{eq:exO8} we infer that the ${\cal O}_8$ contribution is rather small. 
For the inclusive case matters are different as weak annihilation, by which we mean 
contributions from $4$-Fermi operators, is suppressed by powers of $m_b$ in the OPE 
such that  ${\cal O}_8$ might be the leading effect. The latter picture is consistent with the theoretical and experimental findings quoted above.

\section*{Acknowledgements}

We are grateful to Guido Bell, Vladimir Braun,  Thorsten Feldmann, Sebastian J\"ager, Andreas J\"uttner,  Mikolai Misiak, 
Matthias Neubert, Douglas Ross, Christopher Sachrajda and Christopher Smith for discussions on various aspects at various stages of the project.
RZ gratefully acknowledges the support of an advanced STFC fellowship.

\appendix
\setcounter{equation}{0}
\renewcommand{\theequation}{A.\arabic{equation}}

\section{LC-OPE results of the correlation function $\Pi^{V,P}$}
\label{app:results}

Below we present the results of the LC-OPE for the correlation functions
for the vector and pseudoscalar cases using the decompositions 
in Eq.~ \eqref{eq:CFwith_p}. 
We shall use the same decomposition as in Eq.~\eqref{eq:Gdec}, 
$$
g_i^{(s)} = g_i^{(\perp)} + g_i^{(\parallel)} + .. \;, \qquad i = 0..3 \;, \quad 
g_T^{(s)} =  g_T^{(P)}
$$
for the various contributions on the DA \eqref{eq:DA} parts. The dots stand for higher twist contributions 
such as the photon DA discussed in the next appendix.
In order to present our results in a compact way we introduce the following 
abbreviations for the PV-functions:
\begin{alignat}{2}
\label{eq:PVfcts}
& B_a = B_0(u (p_B^2\!-\!P^2),0,m_b^2) \;, \quad & &  B_b =  B_0(p_B^2-P^2,0,m_b^2)   \;, \nonumber \\[0.1cm]
& B_c =  B_0(u p_B^2 \!+\! \bar u q^2 ,0,m_b^2)  \;,  & &  B_d = B_0(p_B^2,0,m_b^2) \;, \nonumber  \\[0.1cm]
& C_a = C_0(p_B^2,u (p_B^2\!-\!P^2),\bar u P^2 \!+\!u q^2 ,0,m_b^2,0)  \;, \quad & &  C_b =  
C_0(p_B^2,p_B^2-P^2,q^2,0,m_b^2,0) \;, \nonumber  \\[0.1cm]
& C_c = C_0(u p_B^2 +\bar u q^2 ,u (p_B^2\!-\!P^2),q^2,m_b^2,0,m_b^2) \;, \quad & &  C_d = C_0(p_B^2,p_B^2-P^2,q^2,m_b^2,0,m_b^2)
\end{alignat}
Note we have only listed the PV-functions which depend on $p_B^2$ as the other ones do not enter
the dispersion representation. Moreover the function on the right correspond to the functions 
on the left at $u = 1$.

\subsubsection*{$V_\perp$-transverse}

We find that for the transverse parts the Lorentz-projections satisfy:
\begin{equation}
\label{eq:results_perp} 
g^{(\perp)}_2 = (1 - q^2/P^2) g^{(\perp)}_3  \;, \quad  g^{(\perp)}_2=(1 - q^2/P^2)g^{(\perp)}_1 \;, \quad g^{(\perp)}_0=0 \;.
\end{equation}
The second relation is a LEET \cite{LEET98,BFS01} relation. 
 It  can be explained
in a straighforward way at the level of the  $\phi_\perp$-distribution in use.
We may factor out the perpendicular $K^*$ DA from the amplitude ${\cal A}^{*\mu}(V)$ to give,
\begin{equation}{\cal A}^{*\mu}(V) = {\rm Tr}\{\s \eta\s p I^\mu\} + \dots\label{eq:kstar-perp-expansion} \;,
\end{equation}
since  the projector is propotional to $\s{p}\s{\eta^*}$  \eqref{eq:DA}.
The dots stand for contributions from other terms in the $K^*$ light-cone expansion.
$I^\mu$ may generally be written as
\begin{equation}
I^\mu(V) = \left[I_0^\mu + I_1\s p\gamma^\mu + I_2\s q\gamma^\mu + I_3\s p\s q\right](1-\gamma_5)
\end{equation}
where terms with an odd number of $\gamma$ matrices have been excluded because they do not contribute to \eqref{eq:kstar-perp-expansion}.
Inserting this form into \eqref{eq:kstar-perp-expansion} then gives
\begin{equation}
{\rm Tr}\{\s \eta\s p I^\mu\} = I_2{\rm Tr}\{\s\eta\s p\s q\gamma^\mu(1-\gamma_5)\}\;,
\end{equation}
and hence there is only a single scalar amplitude which contributes to the result.
Evaluating the trace in our basis \eqref{eq:Vprojectors} yields the identity
\begin{equation}
G_2^{(\perp)}(q^2) + \frac{q^2}{m_B^2}G_3^{(\perp)}(q^2) = G_1^{(\perp)}(q^2)\;.
\end{equation}
As previously noted $G_2^{(\perp)}(q^2)=(1-q^2/m_B^2)G_3^{(\perp)}(q^2)$ so it follows that $G_1^{(\perp)}(q^2)=G_3^{(\perp)}(q^2)$ which shows consistency between the two Eqs. in \ref{eq:results_perp}.
The first relation is of a more general type 
which we would like to explain below:
Decomposing the following matrix element, 
\begin{equation}
\matel{\gamma^*(q,\rho) V(p,\eta)}{H_{\rm eff}} {\bar B(p_B)}  = X_1(q^2) P_1^\rho +
 X_2(q^2) P_2^\rho + X_3(q^2) P_3^\rho \;,
\end{equation}
the following relation must be true
\begin{equation}
X_2 - \left( 1-\frac{q^2}{m_B^2}  \right) X_3 = {\cal O}(m_V) \;,
\end{equation}
in order to cancel an explicit  factor $1/m_V$ in the decay rate \cite{LZ12b}. 
More precisely, by this argument we preclude power divergences which cannot be there
as IR-divergences are at worst logarithmic in four dimensions, as mentioned previously.
Thus, for any projection which does not contain an explicit $m_V$ factor in its definition,
e.g. $\phi_\perp$ but not $\phi_\parallel$, the relation holds up to ${\cal O}(m_V^2)$. E.g. 
$ G_2^{(\perp)} = (1- q^2/m_B^2) G_3^{(\perp)} + {\cal O}(m_V^2)$. Since we neglect $m_V^2$ altogether the first relation in \eqref{eq:results_perp} is a necessary outcome.

We parametrise the result $g_1^{(\perp)}(q^2)$ as
\begin{equation}
\label{eq:gperp}
k_V^{-1}   g^{(\perp)}_1(q^2) = \frac{\alpha_s}{4 \pi} C_F (-\frac{1}{2} ) f^\perp_V m_b^2 Q_b \int_0^1{du} \,  t_H^{(\perp)}(u) \, \phi_\perp(u)
\end{equation}
where the  $t_H^{(\perp)}(u)$  corresponds to the hard kernel and is given
in terms of PV-functions
\begin{equation}
\label{eq:tHperp}
t_H^{(\perp)}(u) = \sum^d_{i = a}( b^\perp_i B_i + c^\perp_i C_i )
\end{equation}
where the sum extends alphabetically from $a$ to $d$. The only non-zero coefficients are
\begin{eqnarray}
\label{eq:bcperp}
 (b^\perp_a, b^\perp_c, b^\perp_d)  &=&    (\frac{q_R}{u q^2 + \bar u P^2},\frac{1}{\bar u q^2 + u P^2}, 
2(b^\perp_a +  b^\perp_c)  )   \;,  \nonumber \\[0.1cm]
(c^\perp_a, c^\perp_c)  &=&  (-2 q_R,-1)  \;,
\end{eqnarray}
with  $q_R \equiv Q_q/Q_b$ being the charge ratio.

\subsubsection*{$V_\parallel$-longitudinal}

The computation of $V_\parallel$ is in principle highly non-trivial due 
the extra coordinates $x$ appearing in front of the integral in \eqref{eq:DA}.
We shall employ though the so-called ultra-relativistic limit,
\begin{equation}
\label{eq:ultra}
\eta(p)_\al \to \frac{1}{m_V} \left(  p_\al  + {\cal O} \left( \frac{m_V^2}{E_V^2} \right) \zeta_\alpha \right) \;,
\end{equation}
which is correct up to the relativistic correction as indicated and the vector $\zeta$ is a linear combination of  $p$ and $\eta$.
In this limit, using the DA as given in appendix \ref{app:DA}, 
the $V_\parallel$ and $P$ contributions are identical up to the replacements
$f_V^\parallel \to - i f_P$ and $\phi_\parallel \to \phi_P$ as can easily be understood by commuting the $\gamma_5$ through the diagram until it is ``annihilated" by $(1+\gamma_5)\gamma_5  = (1+\gamma_5)$ which originates from $\tilde {\cal O}_8$.
Noting that in the ultra-relativistic limit 
\begin{equation}
P_1 \to 0 \;, \quad P_2 \to c P_3|_{\eta \to p/m_V}
\end{equation}
with $c$ a constant it is clear\footnote{The more careful reader might want to know
that $P_2$ corresponds to a term which is  linearly dependent and one that is linearly independent of $P_3$. In the limit the latter vanishes.} that only $g_3$ receives a contribution. 
Taking further into account Eq.\eqref{eq:P3limit} one gets:
\begin{equation}
\label{eq:relation}
G^{(P)}_T(q^2)  =
  \frac{p \cdot Q}{m_V} \frac{i f_P}{f_V^\parallel} \frac{k_V}{k_P} G_3^{(\parallel)}(q^2)|_{\phi_\parallel \to \phi_P} 
= \frac{-(m_B^2 - q^2)}{  2 m_V(m_B-m_P)} \, \frac{ f_P}{f_V^\parallel}  \,
 G_3^{(\parallel)}(q^2)|_{\phi_\parallel \to \phi_P} \;.
\end{equation}
Thus, the result of the longitudinal vector meson entirely follows from the pseudoscalar 
in the ultra-relativistic limit. Note that the sign of this relation changes when 
$(1+\gamma_5) \to (1-\gamma_5)$ in ${\cal O}_8$ \eqref{eq:O8} which is reflected in 
\eqref{eq:prime} as well.

\subsubsection*{$P$ (pseudoscalar)}

Analogous to \eqref{eq:gperp} we parametrise $g^{(P)}_T$ as follows:
\begin{equation}
k_P^{-1}     g_T^{(P)}(q^2) = \frac{\alpha_s}{4 \pi} C_F (-\frac{1}{2} ) f_P m_b^2 Q_b \int_0^1{du} \, t_H^{(P)}(u) \, \phi_P(u) \;.
\end{equation}
The entire expression of $ t_H^{(P)}(u)$ is rather bulky so we shall give only 
one coefficient for  $t_H^{(P)}(u)$,  
\begin{equation}
c_b^P  =  \frac{4 q_R  P^2
   \left(m_b^4+m_b^2 (P^2-2
   p_B^2+q^2)+p_B^2
   (p_B^2-P^2)\right)}{m_b (m_B-m_P) \bar u 
   \left(P^4+2 P^2 q^2+q^2
   (q^2-4 p_B^2) \right)}
    \;,
\end{equation}
which at least allows our results to be verified partially.

 \begin{table}[H]
 \footnotesize
\center

\begin{tabular}{c|cccc}
$ Q^2 $
& $ G_1^{(\perp)}\times 10^2 $
& $ G_3^{(\parallel)}\times 10^2 $
& $ G_1^{(\perp)}\times 10^2 $
& $ G_3^{(\parallel)}\times 10^2 $
\\ \hline
& $ B^-\rightarrow K^{*-} $
& $ B^-\rightarrow K^{*-} $
& $ \bar B^0\rightarrow\bar K^{*0} $
& $ \bar B^0\rightarrow\bar K^{*0} $
\\ \hline
 $ 0.010 $
& $ 0.2931-0.3960i $
& $ 2.3443+0.8303i $
& $ 0.2022+0.1980i $
& $ -1.1952-0.4151i $
\\ $ 0.261 $
& $ 0.3204-0.3781i $
& $ 1.0673+0.8213i $
& $ 0.1661+0.1890i $
& $ -0.5574-0.4107i $
\\ $ 0.512 $
& $ 0.3384-0.3604i $
& $ 0.7999+0.8122i $
& $ 0.1431+0.1802i $
& $ -0.4243-0.4061i $
\\ $ 0.764 $
& $ 0.3526-0.3429i $
& $ 0.6388+0.8029i $
& $ 0.1251+0.1715i $
& $ -0.3443-0.4014i $
\\ $ 1.015 $
& $ 0.3641-0.3257i $
& $ 0.5217+0.7933i $
& $ 0.1102+0.1629i $
& $ -0.2863-0.3966i $
\\ $ 1.266 $
& $ 0.3736-0.3087i $
& $ 0.4286+0.7834i $
& $ 0.0975+0.1544i $
& $ -0.2403-0.3917i $
\\ $ 1.517 $
& $ 0.3815-0.2920i $
& $ 0.3508+0.7732i $
& $ 0.0866+0.1460i $
& $ -0.2020-0.3866i $
\\ $ 1.768 $
& $ 0.3878-0.2755i $
& $ 0.2834+0.7628i $
& $ 0.0771+0.1378i $
& $ -0.1688-0.3814i $
\\ $ 2.020 $
& $ 0.3929-0.2593i $
& $ 0.2235+0.7519i $
& $ 0.0689+0.1297i $
& $ -0.1395-0.3760i $
\\ $ 2.271 $
& $ 0.3969-0.2434i $
& $ 0.1693+0.7407i $
& $ 0.0617+0.1217i $
& $ -0.1129-0.3703i $
\\ $ 2.522 $
& $ 0.3998-0.2277i $
& $ 0.1196+0.7290i $
& $ 0.0554+0.1139i $
& $ -0.0887-0.3645i $
\\ $ 2.773 $
& $ 0.4018-0.2124i $
& $ 0.0734+0.7168i $
& $ 0.0499+0.1062i $
& $ -0.0662-0.3584i $
\\ $ 3.024 $
& $ 0.4028-0.1974i $
& $ 0.0300+0.7041i $
& $ 0.0453+0.0987i $
& $ -0.0451-0.3521i $
\\ $ 3.275 $
& $ 0.4030-0.1827i $
& $ -0.0110+0.6908i $
& $ 0.0413+0.0913i $
& $ -0.0253-0.3454i $
\\ $ 3.527 $
& $ 0.4024-0.1683i $
& $ -0.0500+0.6768i $
& $ 0.0379+0.0842i $
& $ -0.0064-0.3384i $
\\ $ 4.786 $
& $ 0.3883-0.1024i $
& $ -0.2248+0.5935i $
& $ 0.0295+0.0512i $
& $ 0.0775-0.2967i $
\\ $ 6.046 $
& $ 0.3586-0.0489i $
& $ -0.3754+0.4758i $
& $ 0.0323+0.0245i $
& $ 0.1488-0.2379i $
\\ $ 7.305 $
& $ 0.3177-0.0129i $
& $ -0.4908+0.2946i $
& $ 0.0431+0.0065i $
& $ 0.2019-0.1473i $
\\ $ 8.565 $
& $ 0.2758+0.0000i $
& $ -0.4519+0.0224i $
& $ 0.0562-0.0000i $
& $ 0.1770-0.0112i $
\\ $ 9.824 $
& $ 0.2492+0.0000i $
& $ -0.2972+0.0000i $
& $ 0.0630-0.0000i $
& $ 0.0933-0.0000i $
\\ $ 11.084 $
& $ 0.2312-0.0000i $
& $ -0.2485-0.0000i $
& $ 0.0669+0.0000i $
& $ 0.0613+0.0000i $
\\ $ 12.343 $
& $ 0.2176-0.0000i $
& $ -0.2243-0.0000i $
& $ 0.0696+0.0000i $
& $ 0.0400+0.0000i $
\\ $ 13.603 $
& $ 0.2070-0.0000i $
& $ -0.2128+0.0000i $
& $ 0.0718+0.0000i $
& $ 0.0230-0.0000i $
\\ $ 14.862 $
& $ 0.1986+0.0000i $
& $ -0.2101+0.0000i $
& $ 0.0740-0.0000i $
& $ 0.0076-0.0000i $
\\ $ 16.122 $
& $ 0.1921-0.0000i $
& $ -0.2147-0.0000i $
& $ 0.0763+0.0000i $
& $ -0.0080+0.0000i $
\\ $ 17.381 $
& $ 0.1873-0.0000i $
& $ -0.2267+0.0000i $
& $ 0.0790+0.0000i $
& $ -0.0252-0.0000i $
\\ $ 18.641 $
& $ 0.1843+0.0000i $
& $ -0.2475+0.0000i $
& $ 0.0824-0.0000i $
& $ -0.0459-0.0000i $
\\ $ 19.900 $
& $ 0.1831-0.0000i $
& $ -0.2803+0.0000i $
& $ 0.0869+0.0000i $
& $ -0.0725-0.0000i $
\\ $ 21.160 $
& $ 0.1844-0.0000i $
& $ -0.3310-0.0000i $
& $ 0.0932+0.0000i $
& $ -0.1097+0.0000i $
\\ \hline
& $ D^0\rightarrow\rho^0 $
& $ D^0\rightarrow\rho^0 $
& $ D^+\rightarrow\rho^+ $
& $ D^+\rightarrow\rho^+ $
\\ \hline
 $ 0.010 $
& $ -7.0027-4.9787i $
& $ 14.939+2.507i $
& $ -1.9295+2.4893i $
& $ 19.589-1.254i $
\\ $ 0.048 $
& $ -6.5207-4.7048i $
& $ 10.506+2.673i $
& $ -1.8309+2.3524i $
& $ 0.5204-1.3366i $
\\ $ 0.087 $
& $ -6.2041-4.4945i $
& $ 8.9314+2.7462i $
& $ -1.7662+2.2472i $
& $ -1.0918-1.3731i $
\\ $ 0.125 $
& $ -5.9599-4.3583i $
& $ 7.9497+2.9176i $
& $ -1.7163+2.1792i $
& $ -1.4961-1.4588i $
\\ $ 0.163 $
& $ -5.7571-4.2099i $
& $ 7.2213+3.1650i $
& $ -1.6766+2.1050i $
& $ -1.5870-1.5825i $
\\ $ 0.202 $
& $ -5.5875-4.1273i $
& $ 6.6209+3.4331i $
& $ -1.6417+2.0637i $
& $ -1.5541-1.7166i $
\\ $ 0.240 $
& $ -5.4402-4.0195i $
& $ 6.1014+3.5850i $
& $ -1.6115+2.0098i $
& $ -1.4644-1.7925i $
\\ $ 0.440 $
& $ -4.9159-3.4292i $
& $ 3.7348+4.8267i $
& $ -1.4907+1.7146i $
& $ -0.5866-2.4133i $
\\ $ 0.640 $
& $ -4.6317-3.0393i $
& $ 0.8816+6.7486i $
& $ -1.3979+1.5196i $
& $ 0.8464-3.3743i $
\\ $ 0.840 $
& $ -4.4966-2.6815i $
& $ -3.643+11.193i $
& $ -1.3125+1.3407i $
& $ 3.2215-5.5964i $
\\ $ 1.040 $
& $ -4.4921-2.1974i $
& $ -11.832+18.837i $
& $ -1.2174+1.0987i $
& $ 7.4949-9.4187i $
\\ $ 1.240 $
& $ -4.6038-1.8406i $
& $ -27.338+33.651i $
& $ -1.1080+0.9203i $
& $ 15.490-16.825i $
\\ $ 1.440 $
& $ -4.8063-1.2685i $
& $ -58.743+61.022i $
& $ -0.9925+0.6342i $
& $ 31.511-30.511i $
\end{tabular}
\caption{\small $q^2$-dependance of the $G_1^{(\perp)}(q^2)$- and $G_3^{(\parallel)}(q^2)$-functions 
for the four characteristic cases depending on whether the initial state is $B^-,\bar B^0, D^0$ or $D^+$-type.  The tables can be requested from the authors.}
\label{tab:resultsq2}
 \end{table}

\section{Photon distribution amplitude contributions}
\label{app:photonDA}

In this section we present a brief discussion of the contributions due 
to the photon distribution amplitude. The latter corresponds to the 
LD part of the photon whereas the photon of perturbation theory 
corresponds to the SD contribution. They can be separated in a transparent 
way by the background gauge field technique \cite{BBK02}.

We present results for on-shell photon of the two diagrams shown in Fig.~\ref{fig:photonDA-diagrams}
which constitute corrections to the correlation function in Eq.~\eqref{eq:CF} and its diagrams should be added 
to the series in Fig.~\ref{fig:diaA}. Extending our notation to include the photon DA we obtain:

\begin{figure}[ht]
 \centerline{\includegraphics[width=3.2in]{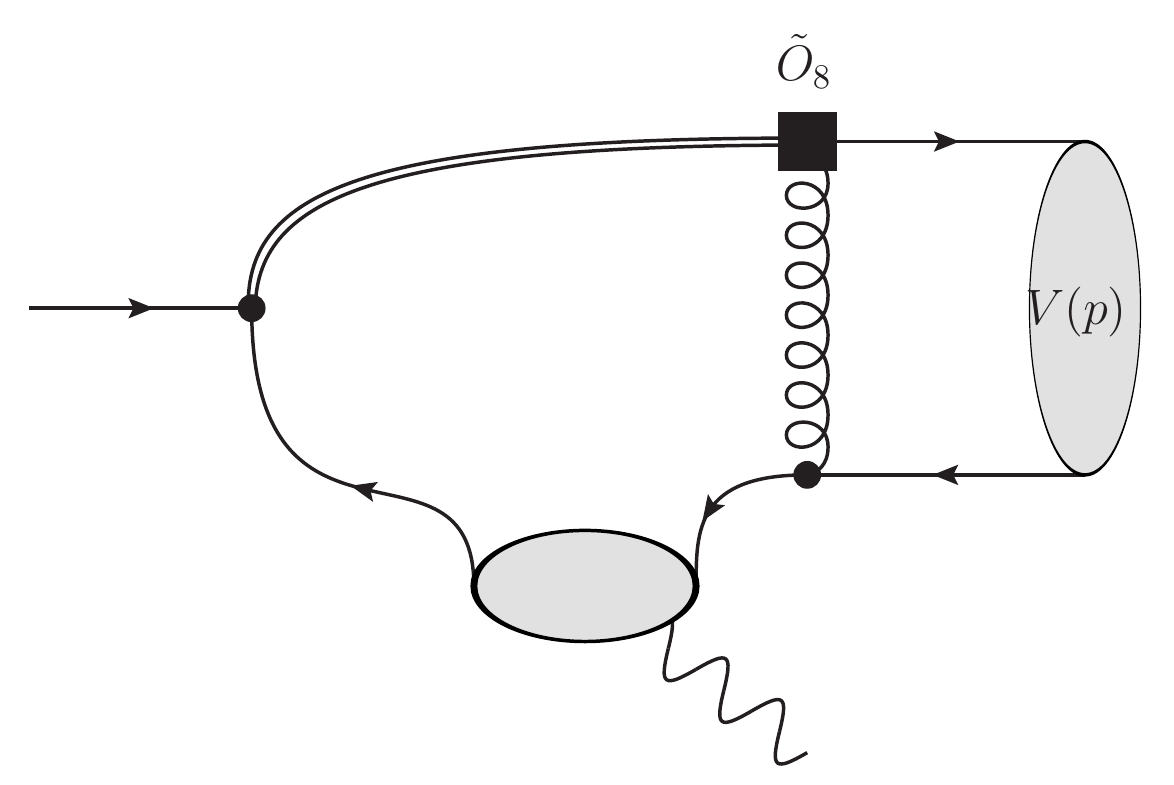}
 \includegraphics[width=3.2in]{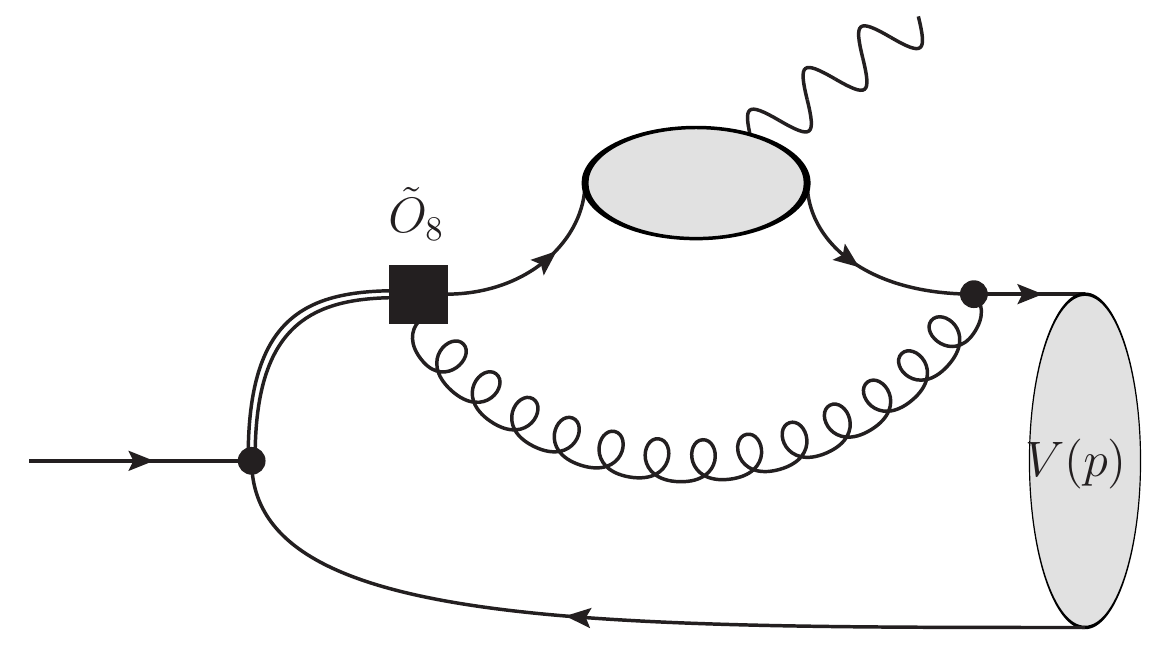}  }
 \caption{\small Additional diagrams arising from the correlation function in Eq.~\eqref{eq:CF}. They correspond to the emission of a LD photon as described in the text.}
 \label{fig:photonDA-diagrams}
 \end{figure}

\begin{equation}
\begin{split}
G_i^{(\phi_\gamma),(\perp,\parallel)}(q^2) =& \frac{f^{(\perp,\parallel)}_{K^*}  \alpha_s m_b C_F }{f_B
m_B^2} \int_{m_b^2}^{s_0} e^{\left(\frac{m_B^2-s}{M^2}\right)}
\Bigg(\chi_s(q^2)\langle\bar ss\rangle \int_0^1
\phi_\perp\left(\frac{m_b^2-q^2}{s-q^2}\right) \phi_\gamma(v)
\rho_i^{s(\perp,\parallel)}(s,v) dv \\
& + \chi_q(q^2)\langle\bar qq\rangle \int_0^1 \phi_\perp(u)
\phi_\gamma\left(\frac{-P^2+q^2+\Delta }{2 q^2}\right)
\theta(m_b^2+P^2-s) \rho_i^{q(\perp,\parallel)}(s,u) du\Bigg) ds
\end{split}
\end{equation}
with 
$
\Delta \equiv \sqrt{(P^2+q^2)^2-4 q^2 (s-m_b^2)}$, 
$ \Omega \equiv \Delta  \left(P^2-q^2\right) (1-2 u)+\Delta$ and $\Sigma \equiv \left(m_b^2-q^2\right)
\left(P^2-q^2\right)+\bar{v} q^2 \left(s-q^2\right)$
and
\begin{alignat}{2}
 \rho^{q\perp}_1 &= \frac{\pi  Q_q \left(-P^2+q^2+\Delta \right)}{6
\Omega} \;, 
& \rho^{s\perp}_1 =& \frac{\pi  q^2 Q_b}{6 \Sigma} \;,  \nonumber \\
 \rho^{q\perp}_2 &= \frac{\pi  Q_q \left(-\left(P^2\right)^2+P^2 \Delta
-q^2 \left(q^2-2 s+\Delta \right)\right)}{6 P^2 \Omega} \;, \quad
&  \rho^{s\perp}_2 =& \frac{\pi  q^2 Q_b \left(q^2-s\right)}{6 P^2
\Sigma}  \;, \nonumber \\
 \rho^{q\perp}_3 &= \frac{\pi  Q_q \left(P^2+q^2-2 s+\Delta \right)}{6
\Omega}  \;,
& \rho^{s\perp}_3 =& -\frac{\pi  Q_b \left(P^2+q^2-s\right)}{6
\Sigma} \nonumber  \;, \\
 \rho^{q\parallel}_3 &= \frac{-2 m_b m_{K^*} P^2 \pi  Q_q}{3 \Omega (m_B^2 - q^2)}  \;, & \rho^{s\parallel}_3 =& \frac{m_b m_{K^*} P^2 \pi  Q_b}{3
\Sigma (m_B^2 - q^2) } \;.
\end{alignat}
The definition of the leading twist-2 photon DA, denoted by $\phi_\gamma(u)$, 
can for example be found in \cite{BBK02}.
Even though the photon DA is of twist-2 and suppressed with regard to the perturbative photon of twist-1,
it is sometimes important because the photon susceptibility $\chi$, somewhat analogous to 
the light meson decay constants, turns out to be rather large e.g. \cite{BBK02}. As it happens though all of 
the expressions above vanish for an on-shell photon $q^2 = 0$, except the $G_3$-part which does though not contribute to the rate at $q^2  =0$. Presumably the vanishing of $G_{1,2}(0)^{(\phi_\gamma),(\perp,\parallel)}$ is accidental and higher twist photon DAs can be expected to contribute. 
One would except the latter to be small though. The extension of the photon DA
to off-shell photon $q^2 >0$ has, to our knowledge, not been discussed systematically in the literature.  One can get an idea of the size of the contributions by using the above computation with $q^2 > 0 $ as well as 
$\chi(q^2)$ of reference \cite{BBK02} in appendix B.  The subscripts $q$ and $s$ for $\chi$ correspond 
to the susceptibility of $ q = u,d$ and an $s$ flavour.
We  find that the contributions are around $5\%$ and thus fairly
negligible in view of the overall uncertainty.

\section{Hadronic input values}
\label{app:input}

The hadronic input for the vector DAs is summarised in Tab.~\ref{tab:input}. 
For  the pseudoscalar decay constants we take $f_\pi = 0.131 \GeV$ and $f_K  = 0.160 \GeV$ \cite{PDG} 
with negligible error and the data for the pseudoscalar meson DAs is  taken from Ref \cite{LT_moments_10}:
\begin{equation}
a_2(\pi) = 0.29(3)(7) \;, \quad a_2(K) = 0.24(3)(7) \;, \quad a_1(K) = 0.074(2)(4)
\end{equation}
The latter value is in good agreement with \cite{SU3reprise}.

The sum rule specific input can be found in Tab.\ref{tab:SRinput}.
We assume $s_0[{f_H}] = s_0[H] \equiv s_0$ throughout. $s_0[B_q] = 35(1) \GeV^2$ is chosen as a reference value. All others are determined 
to satisfy $(m_{H_q}+X)^2 = s_0[H_q]$ for  ``universal" $X$. As discussed previously, $X$ is between
the two pion mass and the rho-threshold. The Borel parameter $M^2[f_H]$ of \eqref{eq:fBSR0} is chosen in the minimum of the Borel window and in addition it is verified that the dimension five operators are below 
$10\%$ and that the
continuum contribution, vulnerable to quark-hadron duality violation, does not exceed $30\%$. 
The Borel parameter $M^2[G]$ for the $G_i$ is chosen such that the continuum is $30\%$; this choice suppresses higher twist-corrections, which we have not computed, maximally. 

\begin{table}
\addtolength{\arraycolsep}{3pt}
\renewcommand{\arraystretch}{1.3}
$$
\begin{array}{c | l l l l l l }
 & f^\parallel [\GeV] & f^\perp [\GeV] & a_2^\parallel  & a_2^\perp & a_1^\parallel & a_1^\perp  \\
 \hline
\rho & 0.216(1)(6) & 0.160(11) & 0.17(7) & 0.14(6)  & - & -  \\
\omega & 0.187(2)(10) & 0.139(18) &  0.15(12) &  0.14(12) & - & - \\  
K^*  & 0.211(7)  & 0.163(8) &  0.16(9) & 0.10(8) & 0.06(4) & 0.04(3) \\
\phi & 0.235(5) & 0.191(6) & 0.23(8) &  0.14(7) & - & - \\
\end{array}
$$
\caption[]{\small Note that $1^{--}$-mesons with odd G-parity have vanishing odd Gegenbauer moments. 
The scale dependent quantities $f^\perp$, $a_{1,2}^{\parallel,\perp}$ are evaluated at $\mu = 1 \GeV$. 
We use the updated value  ${\cal B}(\tau \to K^* \nu_\tau) = 1.20(7)\cdot 10^{-2}$  \cite{PDG} as compared to  the PDG value used by the end of 2006 ${\cal B}(\tau \to K^* \nu_\tau) = 1.29(5) \cdot 10^{-2} $ in \cite{BJZ}, which leads to a decay constant which changes $f^\parallel_{K^*}$ from $0.220\GeV$ to $0.211\GeV$ whereas all the others remain the same as in \cite{BJZ}; 
 with a numerical error corrected for $f_\phi^\parallel$ as noted by the authors of \cite{PichJung}.
The $f^\perp$ decay constants follow from the ratios $r[X] = f^\perp_X(2 \GeV)/f^\parallel_X$ with 
$r[\rho] = 0.687(27)$, $r[K^*] = 0.712(12)$ and $r[\phi] = 0.750(8)$ in \cite{L_fT/t_08}. Further, we use
$r[\omega] \simeq r[\rho]$ in view of a lack of a lattice QCD determination of this quantity.
For the DA parameters we have chosen 
to average $a_1^\parallel$, $a_2^\parallel(\rho,K^*,\phi)$ values from the lattice
\cite{LT_moments_10} with the sum rule determinations keeping the relative sum rule uncertainty, which is larger, in order to account for neglecting higher Gegenbauer moments.
The references for the sum rule values are \cite{Ball:1996tb} for the $\rho$, \cite{BJ07} for the $\phi$ 
and  \cite{SU3reprise} and \cite{BZ06b} for the $K^*$.
In view of the lack of theoretical determinations of parameters for the $\omega$, we have assumed the same
values as for the $\rho$ enlarging the uncertainty by a factor of 2.}
\label{tab:input}
\end{table}

\begin{table}
\addtolength{\arraycolsep}{3pt}\renewcommand{\arraystretch}{1.3}
$$
\begin{array}{l | rrrr | r  || l  r ||  l r  }
H & s_0         & M^2[G] & M^2[f_{H}] & m_H & f_H \eqref{eq:fBSR0}    &  \text{cond.} & \text{value} &  \text{mass} & \text{value}
\\\hline
B_s &  36(1.5)  & 9(2) &  5.0(5)  & 5.37  & 0.162 & \aver{ \bar q  q} &   (-0.24(1))^3   & m_b & 4.7(1)     \\
B_q &  35(1.5)  &  9(2) & 5.0(5) &  5.28 & 0.142 & \aver{ \bar s  s}  &  0.8(1)   \aver{ \bar q  q} & m_c  & 1.3(1)  \\    
D_s &  6.7(7) &  6(2) & 1.5(2)  & 1.96 & 0.185  & \aver{ \bar q G q} & (0.8(1))^2  \aver{ \bar q  q} &  \bar m_s&  0.094(3)  \\
D_q &  6.2(7) & 6(2) & 1.5(2) & 1.86 &   0.156 &   \aver{ \bar s G s} & (0.8(1))^2  \aver{ \bar s   s} &  & 
\end{array}
$$
\addtolength{\arraycolsep}{-3pt}\renewcommand{\arraystretch}{1}
\caption[]{\small (left) $H$ stands for heavy-light meson and $q$ stands for either a $u$ or $d$ quark.
Sum rule specific values in units of $\GeV$ to the appropriate power.  
$f_H$ correspond to the decay constants obtained from 
a tree-level sum rule. They should not be compared with the true value of $f_H$ as the latter have substantial radiative corrections in QCD sum rules.
(middle) condensates relevant for the $f_H$ sum rule \eqref{eq:fBSR0}. 
(right) Quark masses. The tree-level heavy quark masses are chosen to
satisfy  $m_H \simeq m_h + \bar \Lambda$ with $\bar \Lambda \simeq 0.6 \GeV$ approximately. 
The strange quark mass in the $\overline {\text{MS}}$ correspond to $\mu_{\overline {\text{MS}}}  = 2\, \GeV$. In the the sum \eqref{eq:fBSR0} $\bar m_s$ is scaled 
up to $\mu = \mu_F$.}\label{tab:SRinput}
\end{table}

\section{Non-spectator corrections $G^{(ns)}$}
\label{app:non-spectator}

The correction which do not connect the gluon of the operator $\tilde {\cal O}_8$ with the spectator 
quark are depicted in Fig.~\ref{fig:diaA}(bottom). They have been computed for 
the inclusive $b \to s ll$ \cite{AAGW01}.  By gauge invariance the contribution is proportional 
to a function  $F_{8}^{7(9)}(q^2/m_b^2)$ times the operator ${\cal O}_{7(9)}$. The latter reduces to  
the standard  tensor and vector form factors $T_i(f_T)$ and $V,A_i(f_+)$ when taken between 
$B$ and $V(P)$ states.
We find:
\begin{alignat}{2}
\label{eq:Gns}
&G^{(ns)}_i(q^2) &=& \left( -\frac{\alpha_s(m_b)}{4 \pi} \right) \left( \frac{Q_b}{-1/3} \right) \left(F_8^{(7)} \, T_i(q^2) - F_8^{(9)} \frac{q^2}{2 m_b}  \, {\cal V}_i(q^2)  \right) \quad  \;, i = 1..3  \;, \nonumber \\[0.1cm]
&G^{(ns)}_T(q^2) &=&  \left( -\frac{\alpha_s(m_b)}{4 \pi} \right) \left( \frac{Q_b}{-1/3} \right) \Big(F_8^{(7)} \, f_T(q^2) -  F_8^{(9)} \underbrace{ \frac{q^2}{2 m_b} v_T(q^2)}_{- \frac{m_B+m_P}{2 m_b} \, f_+(q^2)} 
 \Big) 
\end{alignat}
where $F_{8}^{7(9)}$ are given in \cite{AAGW01} in terms of an expansion in powers of 
$q^2/m_c^2$ and a logarithm. 
The functions ${\cal V}_i$ and $v_T$ are defined as:
\begin{alignat}{2}
& \matel{V(p,\eta)}{\bar s \gamma^\rho(1\mi\gamma_5) b}{\bar B(p_B)} \; &=& \;  P_1^\rho \, {\cal V}_1 + P_2^\rho \, {\cal V}_2  + P_3^\rho \,  {\cal V}_3   +   [ i (\eta^* \cdot q)q^\rho  ]  {\cal V}_P   \nonumber \\[0.1cm]
& \matel{P(p)}{\bar s \gamma^\rho  b}{\bar B(p_B)} \; &=& \;  P_T^\rho \, v_T  +   q_\rho    v_S 
\end{alignat}
with 
\begin{alignat}{2}
&  {\cal V}_P =  \frac{-2 m_V}{q^2} A_0(q^2)  \qquad \qquad   &{\cal V}_1& =  \frac{-V(q^2)}{m_B+m_V}   \nonumber \\ 
& {\cal V}_2 =    \frac{-A_1(q^2)}{m_B-m_V}  &{\cal V}_3& =  \big( \frac{m_B+m_V}{q^2}    A_1(q^2) -   \frac{m_B-m_V}{q^2}    A_2(q^2) \big) \nonumber \\ 
& v_s =  \frac{m_B^2-m_P^2}{q^2} f_0(q^2)   &v_T& =  \frac{ - (m_B+  m_P)}{q^2}  \, f_+(q^2)   \,,  
\end{alignat}
where $V,A_i,f_+,f_0,f_T,T_i$ are all standard form factor notations in the literature.
Note, as manifested by limiting the sum from $i = 1..3$, the $f_0$($A_0$) component does not contribute
to $B \to Vll$ as the $q^\rho$ vanishes upon contraction with $\bar l \gamma_\rho l$ or the photon polarization 
tensor $\epsilon(q)$.

\section{Lorentz structures}
\label{app:projectors}

The Lorentz structures of the vector meson are given by\footnote{The sign convention for the epsilon tensor is given by 
${\rm tr}[\gamma_5 \gamma_a \gamma_b \gamma_c \gamma_d] = 4i \epsilon_{abcd}$ and
are the ones used in the classic textbook of Bjorken \& Drell.}:
\begin{eqnarray}
\label{eq:Vprojectors}
P_1^\rho &=&  2 \epsilon^{\rho}_{\phantom{x} \alpha \beta \gamma} \eta^{*\alpha} p^{\beta}q^\gamma \nonumber \\
P_2^\rho &=& i \{(m_B^2\mi m_V^2) \eta^{*\rho} \mi 
(\eta^*\!\cdot\! q)(p+p_B)^\rho\} \nonumber \\
P_3^\rho &=&  i(\eta^*\!\cdot\! q)\{q^\rho \mi  \frac{q^2 }{m_B^2\mi m_V^2} (p+p_B)^\rho \}   \;, 
\end{eqnarray}
and the one for the pseudoscalar meson is
\begin{eqnarray}
\label{eq:Pprojectors}
P_T^\rho  &=&  \frac{1}{m_B+m_P}\{(m_B^2-m_P^2) q^{\rho} - q^2 (p+p_B)^\rho\}  
\quad .
\end{eqnarray}
All projectors are transverse, i.e. $q \cdot P = 0$ when on-shell momentum relations
like $p_B^2 = m_B^2$ etc are taken into account.
The structure $P_3 = P_3^\rho \epsilon(q)_\rho$ is absent for an on-shell photon since 
$\epsilon(q) \cdot P_3|_{q^2=0} = 0$ and thus $P_3$ can be seen as a purely longitudinal 
part of the photon.  
Note:
 $P_3^\rho = i/(m_B-m_P)(\eta^*\!\cdot\! q) P_T^\rho|_{m_P \to m_V}$. 

\subsection{Extension to include spurious momentum}
\label{app:projk}

The extension of the Lorentz structures to include the spurious momentum $k$
in the vector case \eqref{eq:Vprojectors} is
\begin{eqnarray}
\label{eq:Vprojk}
(p_1)_\rho  &=&  2 \epsilon^{\rho}_{\phantom{x} \alpha \beta \gamma} \eta^{*\alpha} p^{\beta}Q^\gamma \nonumber  \\[0.1cm]
(p_2)_\rho  &=&  i [( (p_B+p)\cdot Q) \, \eta^*_\rho -(\eta^* \cdot Q) (p_B + p)_\rho] 
 \nonumber  \\[0.1cm]
 (p_3)_\rho  &=&  i [ (\eta^* \cdot Q) Q_\rho -(\eta^* \cdot Q) (p_B + p)_\rho \frac{q^2}{Q \cdot (p_B+p)}] 
 \nonumber  \\[0.1cm]
(p_4)_\rho  &=&  i [ (\eta^* \cdot Q) k_\rho -(\eta^* \cdot Q) (p_B + p)_\rho \frac{k \cdot Q}{Q \cdot (p_B+p)}]   
\end{eqnarray}
and in the pseudoscalar case \eqref{eq:Pprojectors} is:
\begin{eqnarray}
\label{eq:Pprojk}
(p_T)_\rho  &=&  (m_B-m_P)[(  Q_\rho -\frac{q^2}{Q \cdot (p_B+p) } (p_B + p)_\rho]  \nonumber  \\[0.1cm]
(\bar p_{\bar T})_\rho  &=&  (m_B-m_P)[(  k_\rho -\frac{k \cdot Q}{Q \cdot (p_B+p) } (p_B + p)_\rho]  \nonumber \end{eqnarray}
Essentially, we get one more structure due to a linearly independent vector $k$ 
and the projectors are extended such that they remain transverse, i.e. $Q \cdot q = 0$. 
This is easy to verify using $q^2 = Q^2$.
Since $p_3^\rho = (\eta \cdot Q) p_T^\rho$ we have got:
\begin{equation}
\label{eq:P3limit}
p_3^\rho \to \left(  \frac{ip \cdot Q}{m_V(m_B-m_V)} \right)  \, p_T^{\rho} =  \left( \frac{i(P^2 -  q^2)}{2 m_V(m_B-m_V)}  \right) \, p_T^{\rho}  \;,
\end{equation}
in the ultra-relativistic limit $\eta \to p/m_V$ as discussed above and below Eq.\eqref{eq:ultra}.
In the last equality we have used the approximation $p^2 = 0$.

\section{Distribution amplitudes}
\label{app:DA}

The leading twist (twist 2)   DAs for the pseudoscalar (e.g. \cite{K00}) and vector (e.g. \cite{BZBtoV})
mesons are defined as follows,
\begin{eqnarray}
\label{eq:DA}
 \langle K(p) |  [\bar s(x)]_\alpha ..  [q(z)]_\beta | 0\rangle  &= &
 i  \frac{f_K}{4} [\s{p} \gamma_5]_{\beta \alpha}  \int_0^1 \!\! du \,e^{i u x \cdot p + i \bar u  z \cdot p} \,\phi_K(u) + ... \nonumber \\
 \matel{K^*(p,\eta)}{[\bar s(x)]_\alpha .. [q(z)]_\beta }{0} &=& \frac{f^\perp_{K^*}}{4} [\s{\eta}^*(p)\s{p}]_{\beta \alpha} 
\int_0^1 du e^{i u x \cdot p + i \bar u  z \cdot p} \phi_\perp(u)    \\ 
&+& m_{K^*} \frac{f_{K^*}}{4} [\s{p}]_{\beta \alpha}  \frac{\eta^* \cdot (x-z)}{p \cdot (x-z)}
\int_0^1 du e^{i u x \cdot p + i \bar u  z \cdot p} \phi_\parallel(u) + ...   \;, \nonumber
\end{eqnarray}
which we have chosen to be represented by the kaons for definiteness.

\section{Contact terms and Ward-Takahashi identities (WTI)}
\label{app:WTI}

The aim of this appendix is to clarify the issue of non-transverse terms 
in the correlation function $\Pi^{P,V}_\rho$ \eqref{eq:CF}. Let us make two points 
before we draw the conclusion for the significance of the computation 
of the  $G_\iota$-functions.

\begin{enumerate}
\item  We would like to observe that the matrix elements $ {\cal A}^{*\rho}(P,V)$
 are transverse, i.e. $q_\rho{\cal A}^{*\rho}(P,V) =0$,
 by virtue of conservation of the electromagnetic
current $\partial \cdot j^{em} = 0$ or gauge invariance. 
The statement is even true for off-shell photons $q^2 \neq 0$ 
for the SD part defined by 
a current insertion as in Eq.\eqref{eq:matel}.
This is readily  
derived by integration by parts  e.g \cite{Weinberg1}.  Thus we were right to use transverse
projectors only.

\item 
More complicated cases arise from contact terms due to charged
 operator insertions on the level of the  correlation function 
$\Pi_\rho^{P,V}$  \eqref{eq:CF}. 
This is formalised in terms of a WTI-idenity for the correlation function, 
which we have used as a check 
of our computation. 
Consider the correlation function, as depicted in Fig.\ref{fig:Crho},
\begin{figure}[h]
 \centerline{
 \qquad \includegraphics[width=3.5in]{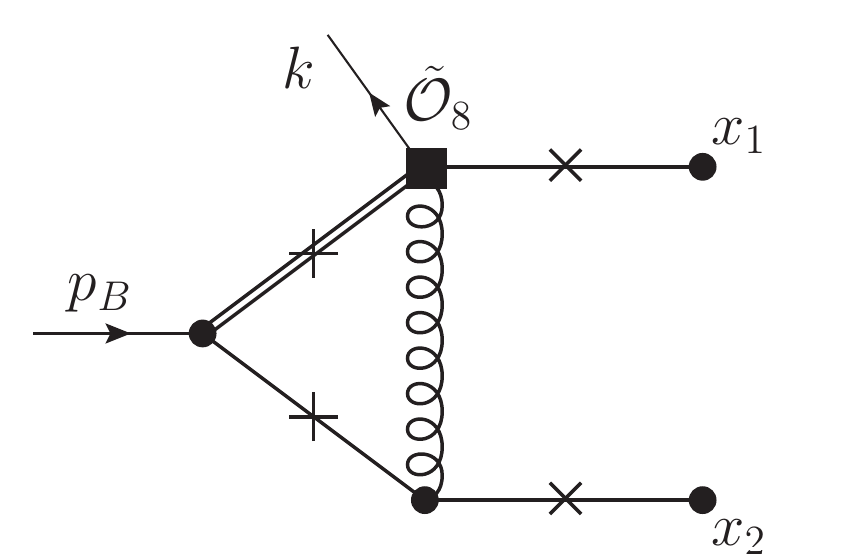}
 }
 \caption{\small Correlation function $C^\rho$ in Eq.~\eqref{eq:CFoff}. The crosses 
 denote  the  four possible places where the perturbative photon of momentum $Q$ can 
 be radiated from.  }
 \label{fig:Crho}
 \end{figure} 
\begin{equation}
\label{eq:CFoff}
C_\rho = i \int_{x,y,z}  \!\!\!\!\!\! e^{-i p_B \cdot x+ i Q \cdot y - i u x_1 \cdot p + \bar u x_2 \cdot p } \matel{0 }
{T J_B(x) \, j_\rho^{\rm em}(y) \, \bar q \s{A} q(z) \,  \bar s(x_1) u \s{p} {\cal P} \bar u  \s{p}  q(x_2)  \tilde O_8(0)}{0} \;,
\end{equation}
with an unspecified projector ${\cal P}$. 
Note, one could equally well leave the two open indices  instead of inserting ${\cal P}$.  This correlation function corresponds to
the one we use in our computation modulo the convolution and the
specific projection ${\cal P}$ of the DA. 
The WTI specifies what happens under contraction with $Q_\rho$:
\begin{equation}
Q^\rho C_\rho = \text{3 contact terms in Fig.\ref{fig:contact}}
\end{equation}
We have verified in each case that this identity is satisfied 
for unspecified ${\cal P}$. The contact terms arise when the derivative acts
on the $T$-product and give rise to  $[j_0, {\cal O}] = q_{\cal O} {\cal O}$-type terms e.g. 
\cite{Weinberg1}, where $q_{\cal O}$ is the charge of the operator ${\cal O}$.
The three contact terms, corresponding to the charged operators, are depicted in Fig.\ref{fig:contact}. 
\begin{figure}[h]
 \centerline{
 \qquad \includegraphics[width=6.0in]{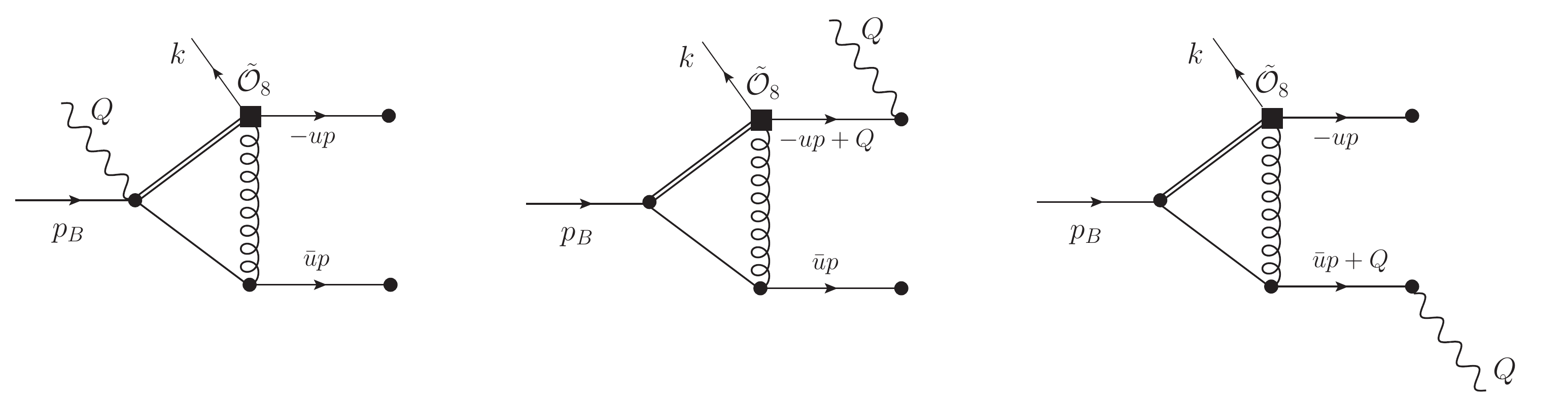}
 }
 \caption{\small Contact terms for the "off-shell" WTI; The diagram on 
 the left is proportional to the charge of the $B$-meson whereas the middle
 and right diagram are proportional to the charge of the $s$-quark and 
 the spectator quark respectively. Only the diagram on the left needs to be computed anew;
 the other two diagrams are proportional to $u p \cdot {\cal A}|_{A_4}$ and  
 $\bar u p \cdot {\cal A}|_{A_1}$ respectively.}
 \label{fig:contact}
 \end{figure}
\end{enumerate}

The question that imposes itself is: how can transversity of the amplitude 
and the non-transversity of the correlation $C_\rho$, used to extract the $G_\iota$-functions, 
be reconciled?
One might think that the contact terms disappear once we  go ``on-shell",
by which we mean specifying  the projector  to be ${\cal P} \sim (\s{p}\gamma_5 , \s{p}, [\s{p}\s{\eta}])$ for the DA $(\phi_P,\phi_\perp, \phi_\parallel)$ respectively.
Non-transverse structures remain for 
for $P/V_\parallel$ but not for $V_\perp$;
$g_0^{(P)} \sim g_0^{(\parallel)} \neq 0$ and $g_0^{(\perp)} = 0$, c.f. Eq.\eqref{eq:results_perp} for the latter.
It is the  diagram to the left, in which the photon is radiated from the charged  $J_{B^-}$, that gives 
a non-vanishing contribution.  The momentum flowing into this vertex is 
$(p_B - Q)^2 = (p + k)^2 = p_B^2-P^2$. The transverse part is  proportional 
to PV-functions of the type $B_0(p_B^2-P^2,0,m_b^2)$ as expected 
and  displays a cut in $p_B^2 > m_b^2 +P^2 = m_b^2 + m_B^2$. 
This contribution can be seen, as yet another, parasitic cut.
It is though of no relevance in the final dispersion integral in $p_B^2$ since the
is well above the continuum 
threshold $s_H \simeq s_0$ in relations like \eqref{eq:exact} and \eqref{eq:SR2}.

\section{Analytic structure and dispersion representation}
\label{app:dispersion}

Let us parametrise a dispersion representation as follows:
\begin{equation}
\label{eq:dispara}
f(p_B^2) = \int_{0}^\infty \frac{\rho_f}{s-p_B^2-i0} + [f(p_B^2)]_{An} + \text{subtractions} \;.
\end{equation}
The polynomial subtraction terms, as previously emphasised,  are of no importance as they 
vanish under the Borel-transformation. 
 The term $[f]_{An} $ corresponds to an anomalous threshold.
Amongst the PV-functions \eqref{eq:PVfcts} present in the results, given in appendix \ref{app:results},
solely  $C_a$\footnote{$C_b$ corresponds to $C_a|_{u \to 1}$ and so we shall not discuss it separately as well as all other functions on the RHS of the list in Eq.~\eqref{eq:PVfcts}} includes an anomalous threshold which
extends into the lower complex half plane, c.f. Fig.~\ref{fig:as},
at physical momenta $P^2, q^2 >0$. This is discussed in section \ref{app:analytic} from various viewpoints.
In addition, the density $\rho_{C_a}$ necessitates many case distinctions, which is not uncommon 
for vertex function e.g. \cite{complex_sing}.

We have checked the dispersion relations by comparing them against LoopTools \cite{LoopTools} which allow
for numerical evaluation of the scalar PV-functions.
Below, we shall quote the results, starting with the anomalous part of $C_a$: 
\begin{equation}
\label{eq:Ca_An}
[C_a(p_B^2)]_{An} =  - 2\pi i\int_{s_+}^{{\rm Re}\,s_+} \frac{ds}{s-p_B^2}\frac{1}{\sqrt\lambda} \;.
\end{equation}
$s_+$ is one of the two solutions of the leading Landau equations of the graph
\begin{equation}
s_\pm = \frac{(1+u)m_b^2 + uP^2 \pm \sqrt{(uP^2 - \bar u m_b^2)^2 - 4u^2m_b^2q^2 - i0}}{2u} \;.
\end{equation}
where the $-i0$ implies that $\Im\,s_+\le 0$.
The densities $\rho_f$ of the representation \eqref{eq:dispara} are:
\begin{eqnarray}
\label{eq:rho}
\rho_{B_a}&=&\left(1-\frac{m_b^2}{u(s-P^2)}\right)\Theta\left(s-\frac{m_b^2}{u}-P^2\right) \nonumber \\
\rho_{B_c}&=&\left(1-\frac{m_b^2}{us+\bar{u}q^2}\right)\Theta\left(s-\frac{m_b^2-\bar{u}q^2}{u}\right) \nonumber \\
\nonumber \rho_{C_a}&=&  \left( \frac{{\rm Im}[C_a]}{\pi} +  \frac{1}{\sqrt\lambda}\left(\log_L \left(\frac{z_+ - z_L}{z_- - z_L}\right) - \log_-\left(\frac{z_+ - 1}{z_- - 1}\right)\right)
  \right) \Theta(s-m_b^2) \\ 
\nonumber \rho_{C_c}&=&\frac{\log\left(\frac{A-\sqrt{\lambda_1 \lambda_3}}{A+\sqrt{\lambda_1 \lambda_3}}\right)}{\sqrt{\lambda_3}}\left[  \Theta\left(s-\frac{m_b^2-\bar{u}q^2}{u}\right) - \Theta\left(s-\frac{m_b^2}{u}-P^2\right)\right] \nonumber \\[0.1cm]
&+& \frac{\log\left( \left(\frac{B-\sqrt{\lambda_2 \lambda_3}}{B+\sqrt{\lambda_2 \lambda_3}}\right)\left(\frac{A-\sqrt{\lambda_1 \lambda_3}}{A+\sqrt{\lambda_1 \lambda_3}}\right)\right)}{\sqrt{\lambda_3}}\Theta\left(s-\frac{m_b^2}{u}-P^2\right)\;,
\end{eqnarray}
where 
\begin{eqnarray}
\nonumber A&\equiv&2 m_b^2 q^2-u \left(q^2-P^2\right) \left(m_b^2+\bar{u}q^2+u s \right)\\
\nonumber B&\equiv&u \left(\left(q^2-P^2\right) \left(m_b^2+u \left(s-P^2\right)\right)-2 q^2 \left(s-P^2\right)\right)\\
\nonumber \lambda_1&\equiv&\lambda\left(us+\bar{u}q^2,m_b^2,0\right) \;, \quad
\nonumber \lambda_2\equiv\lambda\left(u(s-P^2),m_b^2,0\right) \;, \\
\lambda_3&\equiv& \lambda\left(us+\bar{u}q^2,u(s-P^2),q^2\right)  \;, \quad 
\lambda \equiv \lambda(p_B^2,\bar u P^2+uq^2, u(p_B^2-P^2))
\end{eqnarray}
and $\lambda(x,y,z) = (x- (y+z))^2 - 4 y z$ is the K\"all\'{e}n-function.

The notation $\log_-$ and $\log_L$ in the density $\rho_{C_a}$ demands clarification:
\begin{equation}
\log_L \theta \rightarrow
\begin{cases}
r_+ > 0 \wedge r_- > 0 & \log_+\theta \\
r_+ < 0 \wedge r_- > 0 &
\begin{cases}
\lambda < 0 &
\begin{cases}
s < \ReL  s_+ & \log_+\theta \\
s > \ReL  s_+ & \log_-\theta
\end{cases} \\
\lambda > 0 &
\begin{cases}
\theta < 0 & \begin{cases}
s < \lambda_- & \log_-\theta\\
s > \lambda_+ & \log_+\theta
\end{cases} \\
\theta > 0 & \begin{cases}
\ReL  s_+ < s < \lambda_- & \log\theta - 2\pi i \\
\lambda_+ < s < \ReL s_+ & \log\theta + 2\pi i \\
\text{otherwise} & \log\theta
\end{cases}
\end{cases}
\end{cases} \\
r_+ < 0 \wedge r_- < 0 & \log_-\theta
\end{cases}
\end{equation}
The square root of $\lambda$, but not $\lambda_{1,2,3}$, in Eq.~\eqref{eq:rho} is
to be taken as:
\begin{equation}
\sqrt\lambda \rightarrow
\begin{cases}
\sqrt\lambda & s < \lambda_- \\
i\sqrt{-\lambda} & \lambda_- < s < \lambda_+ \\
-\sqrt\lambda & s > \lambda_+  
\end{cases} \;.
\end{equation}
Furthermore, $\log_\pm$ are defined as follows:
\begin{align}
\log_+ x &= \begin{cases}\log x & \ImL  x = 0 \\ \log (-x) + i\pi & \ImL  x \neq 0\end{cases} \\
\log_- x &= \log(-x) - i\pi
\end{align}
The remaining variables in $\rho_{C_a}$ are given by:
\begin{eqnarray}
\lambda_\pm &=&  \frac{\bar u P^2 + u(1+u)q^2 \pm 2u\sqrt{q^2(\bar uP^2 + uq^2)}}{\bar u^2}  \;, \quad  \lambda =  \bar u^2 (s-\lambda_+)(s-\lambda_-) 
 \nonumber \\
z_\pm &=& \frac{(1+u)p_B^2 - P^2 - uq^2\pm\sqrt\lambda}{2p_B^2} \;, \quad 
z_L = 1 + \frac{\bar uP^2 + uq^2}{m_b^2-p_B^2} \nonumber \\
r_\pm &=& r(\lambda_\pm) \;, \quad r(p_B^2) =  (1 + u - 2z_L)p_B^2 - P^2 - uq^2 \;.
\label{eq:ca-vars}
\end{eqnarray}

\subsection{Analytic structure of  $C_0(s, s - \beta , \alpha  ,0,m_b^2,0)$ in $\C_s$  }
\label{app:analytic}

In this section we shall discuss the analytic properties of the PV-function $C_a $
through a function with simplified but equivalent variables, namely,
\begin{equation}
\label{eq:toy}
C_0(s, s - \beta , \alpha  ,0,m_b^2,0) \;,
\end{equation}
with conventions as indicated in the caption of Fig.~\ref{fig:as}.
The function \eqref{eq:toy}  corresponds to $C_a$ in Eq.~\eqref{eq:PVfcts} with the following substitutions:
\begin{equation}
\label{eq:trans}
s = p_B^2 \;, \quad \alpha = u q^2 + \bar u P^2 \;, \quad \beta = u P^2 + \bar u s \;.
\end{equation}

It is argued in a succession of rigour: first from the viewpoint of Landau equations \ref{app:Leq}, 
then explicit one-loop solutions \& uniqueness of analytic continuation \ref{app:explCS}
and finally axiomatic results by K\"all\'{e}n \& Wightman \ref{app:KW},
 that the correlation function has a complex anomalous threshold 
 on the physical sheet for 
 \begin{equation}
 \label{eq:alphastar}
 \alpha >  \alpha^* \equiv \frac{\beta^2}{4 m_b^2} \;.
 \end{equation}

\subsubsection{Singularities from the Landau equations}
\label{app:Leq}

The Landau equations \cite{Smatrix,Todorov} are a means to determine singularities 
of a perturbative diagram\footnote{Singularities which arise due to infinite
loop-momentum are possible to interpret through the Landau equations though  not easily 
and have therefore been called singularities of the second-type or non-Landau singularities.}.
The crucial and limiting point is that, unless the singularities are real,  there is no direct way to determine
on which Riemann sheets they appear.

\begin{figure}[h]
\centerline{
\hskip0.2in
\includegraphics[width=3.7in]{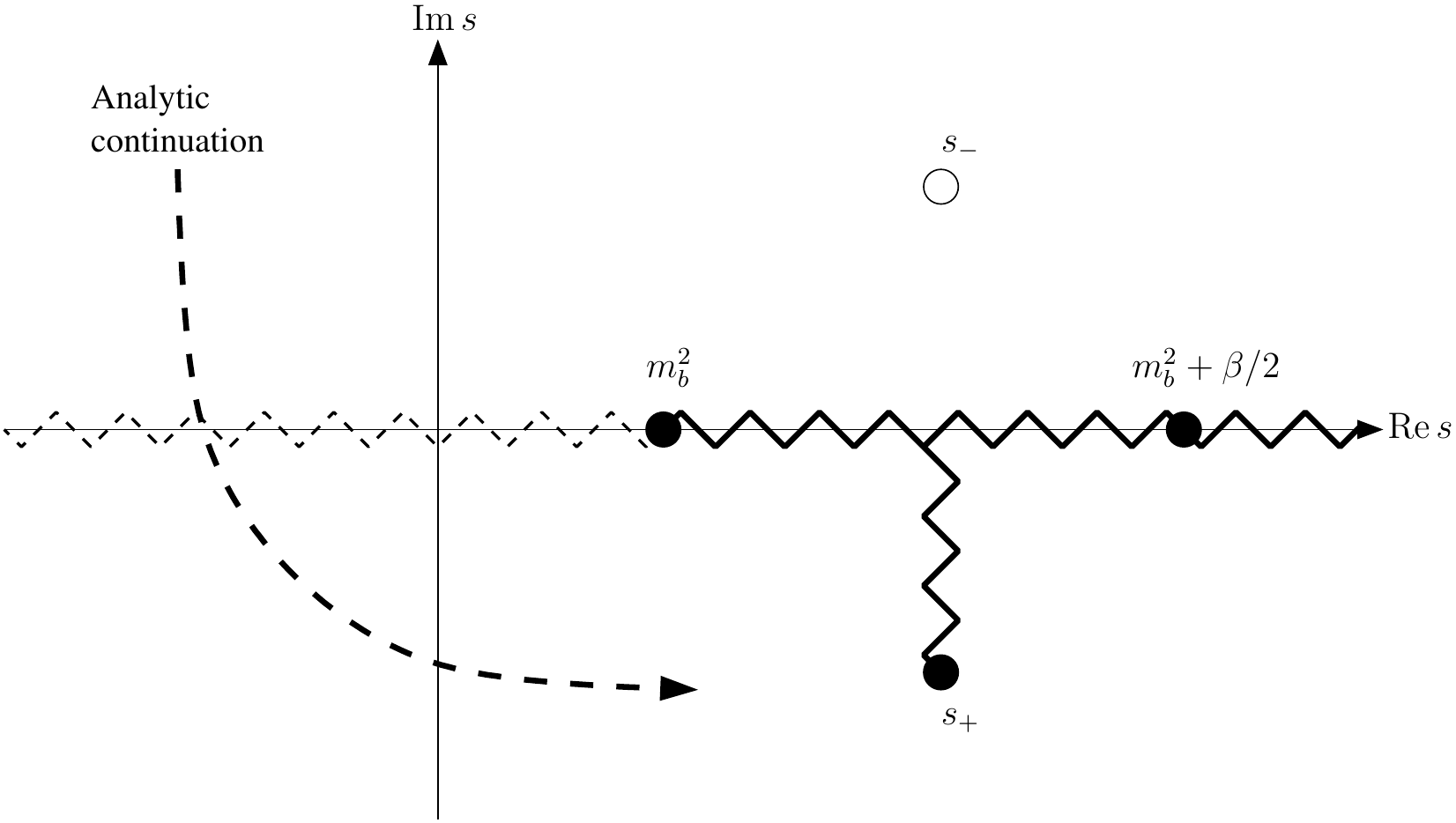}
\hskip0.1in
\includegraphics[width=2.5in]{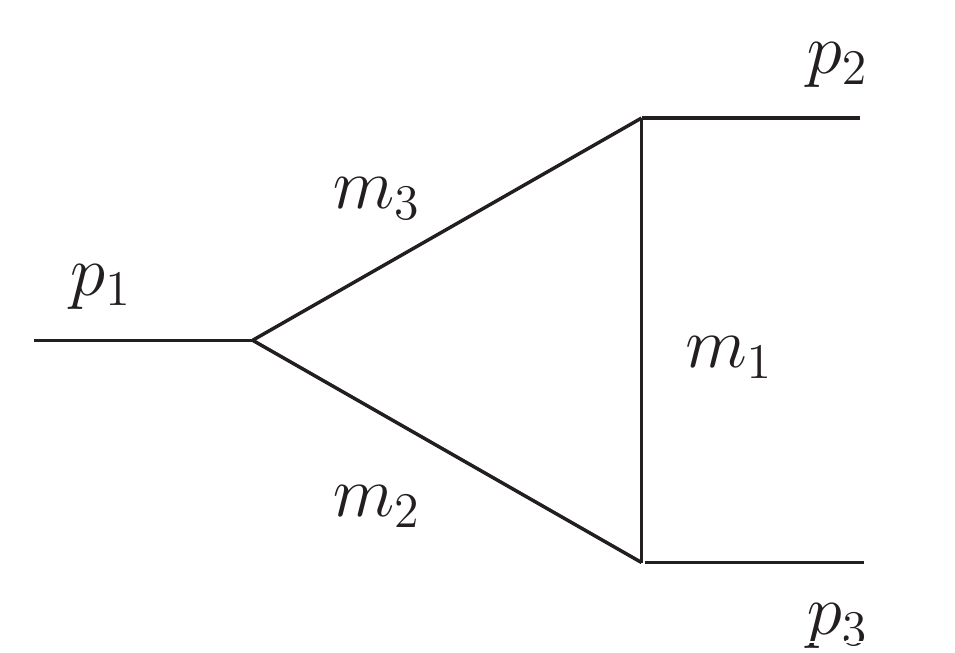}
}
 \caption{\small Analytic structure of  $C_0(s, s - \beta , \alpha  ,0,m_b^2,0)$.  The path of the branch cut connected to the branch point $s_+$ can be inferred from a deformation analysis as in \cite{Mandelstam60}.
 (left) Black spots correspond to branch points on the physical sheet. White spot branch point which is not on 
 the physical sheet. Black zig-zag lines are branch cuts on the physical sheet. The dashed zig-zag line corresponds to a branch cut of $C_a^F$ \eqref{eq:ca-feynman-integral} but not of $C_a = C_0(s, s - \beta , \alpha  ,0,m_b^2,0)$ as explained in the text. The arrow indicates around which branch point $C_a^F$ is analytically 
 continued into the lower half plane. 
 (right) Triangle graph corresponding to the $C_0(p_1^2,p_2^2,p_3^2,m_2^2,m_3^2,m_1^2)$ PV-function. The conventions are the same as in LoopTools \cite{LoopTools} and Feyncalc \cite{Mertig:1990an}.
}
 \label{fig:as}
 \end{figure}

We shall be interested in determining the so-called leading Landau singularity of the 
triangle graph \ref{fig:as}, also known as an \emph{anomalous threshold}. 
It corresponds to all three propagators being on-shell. 
The condition can conveniently be written in terms of a determinant,
\begin{equation}
{\rm det} \begin{pmatrix}
 1 & x_1 & x_2  \\
x_1 & 1  & x_3 \\ 
x_2 & x_3 & 1 
\end{pmatrix}  = 0  \;, \qquad x_i \equiv \frac{p_i^2 - m_j^2 - m_{k}^2}{ 2 m_j m_k} \;, \quad  i \neq j \neq k \neq i \;,
\end{equation}
where $m_j$ and $m_k$ are the masses of the propagators adjacent to the in-going 
momentum squares $p_i^2$.
For the $C_0$ in question \eqref{eq:toy}, this leads to 
the Landau surface
\begin{equation}
\label{eq:Landau_eq}
(s- m_b^2)(s-m_b^2 -\beta) + \alpha m_b^2 = 0
\end{equation}
 whose solutions are given by
 \begin{equation}
 \label{eq:spm}
 s_{\pm} =  m_b^2 + \beta/2  \pm \sqrt{(\beta/2)^2 -  \alpha m_b^2}
 \end{equation}
 As long as $\alpha < \alpha^*$ \eqref{eq:alphastar} the solutions are real 
 and we can decide of whether they are on the physical sheet or not by checking whether the
 Landau equations admit solutions where the Feynman parameter admit values between $[0,1]$.
 As a matter of fact for any $q^2 > 0$, c.f. Eq.~\eqref{eq:trans}, 
 there is exists some $u \in [0,1]$ for which $\alpha > \alpha^*$. Thus we are lead to the question of whether or not the  singularities $s_\pm$
 are on the physical sheet.  
 Some  guidance can be gained following 
  Mandelstam contour deformation prescription \cite{Mandelstam60}.
  The idea is that one starts with values for $P^2$ and $Q^2$ such that $s_\pm$ are real.  Then a dispersion representation can be constructed by checking which singularities are on the physical sheet.
  Upon deformation of  the external momenta ($P^2,Q^2$) 
   the contour is deformed such that no singularities are crossed. 
   Applying this procedure we found that  $s_+$ is on the physical sheet and $s_-$ on an unphysical sheet.
 In the next section we shall show the same result to be true in a more explicit and possibly more 
 transparent way from the known one loop result.

\subsubsection{Complex branch points in the lower half-plane from analytic continuation of the 
Feynaman parameter representation}
\label{app:explCS}

Here we discuss the function $C_a$  \eqref{eq:PVfcts} itself rather than $C_0$ \eqref{eq:toy} because reference is made to the variables used in $\rho_{C_a}$ \eqref{eq:rho} and thereafter. 
Variables are restricted to  the following values :
$0 \leq u \leq 1$, $m_B^2 > m_b^2 >0$, $P^2 = m_B^2 + i0$ and $q^2 -i 0= {\rm Re} [q^2] > 0 $. 
Our two main ingredients are the uniqueness  of analytic continuation  from the real line and the fact that the lowest cut on the real line starts at $m_b^2$.
The latter can be verified from the Landau equations. 

The correlation function $C_a$, originally defined just above the real line of $p_B^2$ (at ${\rm Re}[p_B^2] +i0$), can be  analytically  continued  
into the entire upper half-plane  by the Feynman-parameter integral representation,
\begin{equation}
C_a^F(p_B^2) \! = \!\! \int_0^1 dx \int_0^{1-x} \!\!\! dy \left[(1-x-y)(x p_B^2+yu(p_B^2-P^2)-m_b^2)+xy(\bar u P^2+uq^2) + i0\right]^{-1} \!\! \;,
\label{eq:ca-feynman-integral}
\end{equation}
since it is free from singularities in this region.
For $\Im[p_B^2]\neq 0$ (where the $i0$-prescription is irrelevant) $C_a^F(p_B^{2*})=C_a^F(p_B^2)^*$ by inspection.
This implies that $C_a^F$, but not necessarily $C_a$, has got a branch cut on the real axis whenever 
$\Im [C_a^F(p_B^2)] \neq 0$. Note these are the only possible singularities for the range of variables 
mentioned above.

Using the Feynman-parameter representation $C_a^F(p_B^2)$ as a starting point we construct an analytic continuation to 
the lower half-plane as follows:
\begin{equation}
C_a(p_B^2) = \left\{\begin{matrix}C_a^F(p_B^2) & \Im[ p_B^2 ] > 0\\C_a^F(p_B^{2*})^* + C_a^{\text{rem}}(p_B^2) & \Im [p_B^2] < 0\end{matrix}\right. \;.
\label{eq:ca-split}
\end{equation}
With reminder-function $C_a^{\text{rem}}(p_B^2)$  such that there is no branch cut below 
$p_B^2 < m_b^2$ for $C_a(p_B^2)$.
To remove the branch cut near a given $p_B^2$ we require that $C_a(p_B^2)$ in \eqref{eq:ca-split} is equal immediately above and below the real line which enforces
\begin{equation}
C_a^{\text{rem}}(p_B^2) = 2i\,\Im [C_a^F(p_B^2)] \;, \quad \Im [ p_B^2]=0 \;.
\label{eq:ca-rem-def}
\end{equation}
The resulting function eliminates the branch cut for $p_B^2<m_b^2$.
In this region a remainder function $C_a^{\text{rem}}(p_B^2)$ may be derived from \eqref{eq:ca-rem-def} and \eqref{eq:ca-feynman-integral} using $1/(x+i0)  =  {\rm PP}[ 1/x] - i \pi \delta(x)$\footnote{${\rm PP}$ stands 
for the principal part.}  to give
\begin{equation}
C_a^{\text{rem}}(p_B^2) = -\frac{2\pi i}{\sqrt\lambda}\left(\log\left(\frac{z_+-z_L}{z_+-1}\right) - \log\left(\frac{z_--z_L}{z_--1}\right)\right) \;,
\end{equation}
with $z_\pm$, $z_L$ and $\lambda$ as in \eqref{eq:ca-vars}\footnote{Note whilst the directions of the 
cuts are ambiguous the branch points  $s_\pm$  are unambiguous. Fortunately it is the latter we are interested in.  In other words: The exact location of the cuts is somewhat analogous to the choice of a coordinate system whereas the branch points are not dependent on it.}.
The branch points of the logarithms and square roots  appear on all Riemann sheets unless there are cancellations between terms.

The branch cuts of the two logarithms start at $z_\pm=z_L$ (there are no solutions 
for  $|p_B^2| < \infty $ to $z_\pm=1$), which occurs at $p_B^2=s_\pm$,
and since the branch points $s_\pm$ are separate no cancellation occurs and there indeed must be a cut on all Riemann sheets of $C_a^{\text{rem}}(p_B^2)$.
$s_\pm$ is complex for physical momenta, and since we know that $C_a^{\text{rem}}(p_B^2)$ is the only term with branch points
away from the real line in \eqref{eq:ca-split} we conclude that analytically continuing \eqref{eq:ca-feynman-integral} to $\Im [p_B^2]<0$ across the real line, to the left of the branch point $p_B^2 = m_b^2$ c.f. Fig.~\ref{fig:as}(left), 
necessarily results in a branch cut off the real line in the lower complex half plane.
To this end we note that  $C_a^{\text{rem}}(p_B^2)$ corresponds to $\rho_{C_a}$ \eqref{eq:rho} 
modulo the imaginary part.  To this end we would like to add a clarifying remark. 
Whereas the Feynman parameter representation  does satisfy the Schwarz reflection principle $(C_0^F(s^*))^*  = C_0^F(s)$, as previously stated, the proper analytic continuation $(C_0(s^*))^* \neq C_0(s)$ does not. This is surely due to the complex singularity on the lower half-plane which is not balanced by a singularity on the upper
half plane.

In the next section we are going to learn that the complex singularities are not an artefact of perturbation 
theory but are expected on most general grounds from axiomatic approaches.

 
\subsubsection{The K\"all\'{e}n-Wightman domain}
\label{app:KW}

Based on  axioms such as Lorentz-covariance,  assumption on the 
spectrum and microcausality K\"all\'{e}n \& Wightman \cite{KW58}  obtained 
results on  the domain analyticity of the vacuum expectation value of three scalar fields. 
We note that the $C_0$ PV-function is simply a one-loop approximation in a specific 
theory with three point interactions. 
Denoting the three invariant momentum squares of the three vertices 
by $Z_i = p_i^2$, for $i = 1..3$,  the domain can be separated into eight regions characterised by  
the signs of ${\rm Im}[Z_i]$; denoted by  $[\pm\pm\pm]$ . Those eight octants are partly separated by the normal
cuts. In addition the domains with signatures $[++-]$ and $[--+]$ and permutations thereof 
have got the  following boundaries \cite{Anderson72}:
\begin{equation}
\label{eq:axiomatic}
(Z_1-r)(Z_2-r) +  r Z_3  = 0 \;, \qquad r > 0 \;;
\end{equation}
with ${\rm Im}(Z_1) {\rm Im}( Z_2) > 0 $.
Thus for $(Z_1,Z_2,Z_3) = (s,s-\beta,\alpha+i0)$ with ${\rm Im}[s] <0$ 
we find
\begin{equation}
(s- r)(s -\beta-r) + r \alpha = 0
\end{equation}
which corresponds to the Landau surface equation \eqref{eq:Landau_eq}
upon identifying $r = m_b^2$.

\end{document}